\documentclass[aps,prstper,reprint,groupedaddress]{revtex4-2}
\usepackage{graphicx, tabularx, longtable, nicefrac, amsmath,enumerate, supertabular, eurosym}

\usepackage{booktabs}
\usepackage[latin1]{inputenc}
\usepackage{booktabs}
\usepackage{hyperref,xcolor}

\begin{document}


\title{The impact of an interactive visualization and simulation tool on learning quantum physics: Results of an eye-tracking study}




\author{Stefan K{\"u}chemann$^1$}
\email{mailto: s.kuechemann@lmu.de}
\author{Malte Ubben$^2$}
\author{David Dzsotjan$^3$}
\author{Sergey Mukhametov$^3$}
\author{Carrie A. Weidner$^4$}
\email{mailto: c.weidner@bristol.ac.uk}
\author{Linda Qerimi}
\author{Jochen Kuhn$^1$}
\author{Stefan Heusler$^2$}
\author{Jacob F. Sherson$^5$}

\affiliation{$^1$ Chair of Physics Education Research, Faculty of Physics, Ludwig-Maximilians-University Munich, Geschwister-Scholl-Platz 1, 80539 Munich, Germany}
\affiliation{$^2$ Institute of physics education research, M{\"u}nster university, 48140 M{\"u}nster, Germany}
\affiliation{$^3$ Physics Education Research, Department of Physics, TU Kaiserslautern, Erwin-Schr\"odinger Str. 46, 67663 Kaiserslautern, Germany}
\affiliation{$^4$ Quantum Engineering Technology Laboratories, H. H. Wills Physics Laboratory and Department of Electrical and Electronic Engineering, University of Bristol, Bristol BS8 1FD, United Kingdom}
\affiliation{$^5$ Niels Bohr Institute, University of Copenhagen, 2100 Copenhagen, Denmark}


\date{\today}

\begin{abstract}

Employing scientific practices to obtain and use information is one of the central facets of next generation science standards. Especially in quantum technology education, the ability to employ such practices is an essential skill to foster both academic success and technological development. In order to help educators design effective instructions, the comparison between novices' and experts' eye movements allows the identification of efficient information extraction and integration strategies. In this work, we compare the gaze behavior of experts and novices while solving problems in quantum physics using an interactive simulation tool, Quantum Composer, which displays information via multiple external representations (numerical values, equations, graphs). During two reasoning tasks, we found that metarepresentational competences were crucial for successful engagement with the simulation tool. The analysis of the gaze behavior revealed that visual attention on a graph plays a major role and redundant numerical information is ignored. Furthermore, the total dwell time on relevant and irrelevant areas is predictive for the score in the second task. Therefore, the results demonstrate which difficulties novices encounter when using simulation tools and provides insights for how to design effective instructions in quantum technology education guided by experts' gaze behavior.

\end{abstract}


\maketitle

\section{Introduction}
The understanding of basic quantum principles paved the way for central technological inventions such as transistors, microprocessors, and lasers during the first quantum revolution. Several decades of intensive research efforts initiated the second quantum revolution and now lead to an abundance of novel quantum technologies such as quantum computing, quantum sensing, quantum cryptography and many more. Quantum technology education will be one of the main drivers of commercial and academic gain in the context of the second quantum revolution (\cite{fox2020preparing}. Accordingly, there are several initiatives around the world aimed at teaching principles of quantum physics at high school level as well as in undergraduate and graduate courses. 

Due to the high value of quantum technologies, it is highly relevant to design effective instructions, identify common student difficulties, and to use digital technologies to enhance learning. The goal of an effective instruction is to enable amateurs to reach the competences of experts, and therefore it is a common practice to analyze the problem-solving behavior of experts and to design instructions based on this (\cite{larkin1980expert}. 

In the last decades, digital technologies have taken a pivotal role in education and education research (\cite{scherer2019technology}. Among others, interactive simulation tools offer the opportunity to manipulate parameters and observe the outcome in real time, which has been shown to stimulate students' drive to explore and inquire. However, in such environments students often deal with multiple external representations, each of which has their own of affordances (\cite{fredlund2014unpacking}. Therefore, it is necessary that students possess representational competence to learn in such environments with multiple external representations (\cite{kuchemann2020students}. Previously, in the context of external representations, the analysis of the eye movements of experts in comparison to those of novices has been an effective way to understand efficient problem-solving strategies and to design instructions (\cite{klein2019student}.

In this work, we aim to compare the gaze behavior of experts in quantum physics and novices who have a low level of content knowledge in quantum physics when solving problems in quantum physics using Quantum Composer, an interactive simulation and visualization tool for quantum physics \cite{zaman2021quantum}. Quantum Composer allows the user to manipulate multiple external representations, namely equations, numbers, and graphs, to understand concepts in quantum physics. As the ability to use such simulation tools efficiently is a central aspect for obtaining conceptual understanding, we focus on the following research questions: 
\begin{enumerate}
    \item How successfully do experts and novices use Quantum Composer to solve quantum physics problems? 
    \item How do experts and novices visually process information when solving quantum physics problems using Quantum Composer?
\end{enumerate}

\section{Background}
\subsection{Teaching quantum physics}

In recent years we have been witnessing the emergence of a ``quantum industry"  where the commercialization of quantum technologies like quantum sensing, quantum communication, quantum computing and others is becoming more and more prominent (Heather,2020). In order to fulfill the growing demand for capable experts in this field, the necessary knowledge and skills must be imparted to students during their time in higher education. Thus, to meet these challenges, the community needs to revise and update the ways quantum physics is taught \cite{Carrie_2020}. 

An overarching initiative for creating an educational ecosystem in the field of quantum physics is the European Competence Framework for Quantum Technologies \cite{QTEdu}. The framework, which maps out the landscape of concepts and competences in quantum technologies, aims to be a starting point for planning and structuring educational as well as training projects in this field.

The structure of the framework was inspired by \cite{DigCompEdu}, the Digital Competence Framework for Educators, a successful initiative to create a coordinated, connected educational ecosystem in the field of digital technologies.

In the process of compiling the competence framework, QTEdu used a bottom-up approach where first a three-round Delphi study was conducted with many participants from the quantum technology community, followed by expert interviews for each competence area in order to refine the results \cite{gehrke2020experiencing}. On the long term, the goal of the initiative is an adaptable European quantum education community where stakeholders can have a seamless synergy. Within the initiative, the competence framework's main objective is to provide a frame of reference and establish a common language.
The competence framework contains eight categories that cover (1) concepts of quantum physics, (2) physical foundations of quantum technologies, (3) enabling technologies, (4) hardware for quantum computers and sensors, (5) quantum computing and simulation, (6) quantum sensors and metrology, (7) quantum communication (8) practical and soft skills. These competence categories are not only meant to be a guide for the development of school and university curricula, but also to specify job profiles of industry professionals. 

The basic concepts tested in this study can be localised within the competence framework as superposition and interference; statistical nature of quantum physics; unitary evolution and tunneling. 
Therefore, we address the basis concepts of quantum physics (1.1) and a minor aspect of quantum programming languages and tools (5.2). 

Students' grip on these concepts has been extensively researched in the recent past. Studies on visual \cite{passante2019enhancing} and general \cite{emigh2015student} understanding of time dependence in quantum systems, or on the learning of quantum tunneling \cite{mckagan2008deeper, wittmann2005addressing,morgan2006examining} suggest that students often struggle with grasping these vital concepts in quantum mechanics. Therefore, there is a very real need for teaching methods that facilitate a better, deeper understanding of these topics.

In a parallel initiative, Fox and colleagues conducted series of interviews within 21 companies that develop quantum technologies \cite{fox2020preparing}. The authors found that there are skills in the context of quantum technologies such as coding, statistics and data analysis, troubleshooting, and the understanding of decoherence and noise mechanisms that are relevant across the quantum industry.

Quantum physics in today's higher education is generally taught based on textbooks like \cite{Griffith_1995, Cohen-Tannoudji_1977, Shankar_2011}, which give a great theoretical background and provide information for a deep understanding of the covered concepts. However, with the growing prominence of the quantum industry comes also the need for a more application-oriented, hands-on approach with a broader inclusion that just physics majors, an approach for which the current higher education is not necessarily well equipped.  In order to meet the new challenges, the community needs to create new quantum learning material at all levels \cite{stadermann2019analysis,aiello2021achieving}, novel, efficient teaching methods \cite{singh2015review, krijtenburg2017insights}, and tools to visualize and explore quantum systems (QUILTS,QUVIS,PhET.QM)

Quantum Composer ties into this effort as a visualization/exploration tool that, through simulation, helps students reason and determine correct answers to problems posed in simple quantum systems \cite{Carrie_2020}. Studies have shown that visualization and simulation tools are a great way to assist students in understanding quantum mechanics. Through their use, students can develop scientific understanding by making connections with pre-existing knowledge \cite{cataloglu2002testing}, building mental models \cite{kohnle2018quvis}, as well as gaining insight by developing a visual understanding of quantum physics concepts \cite{passante2019enhancing, kohnle2010developing}.

\subsection{Simulation tools: Quantum Composer}
\begin{figure*}[t!]
\centering
\includegraphics[width=\textwidth]{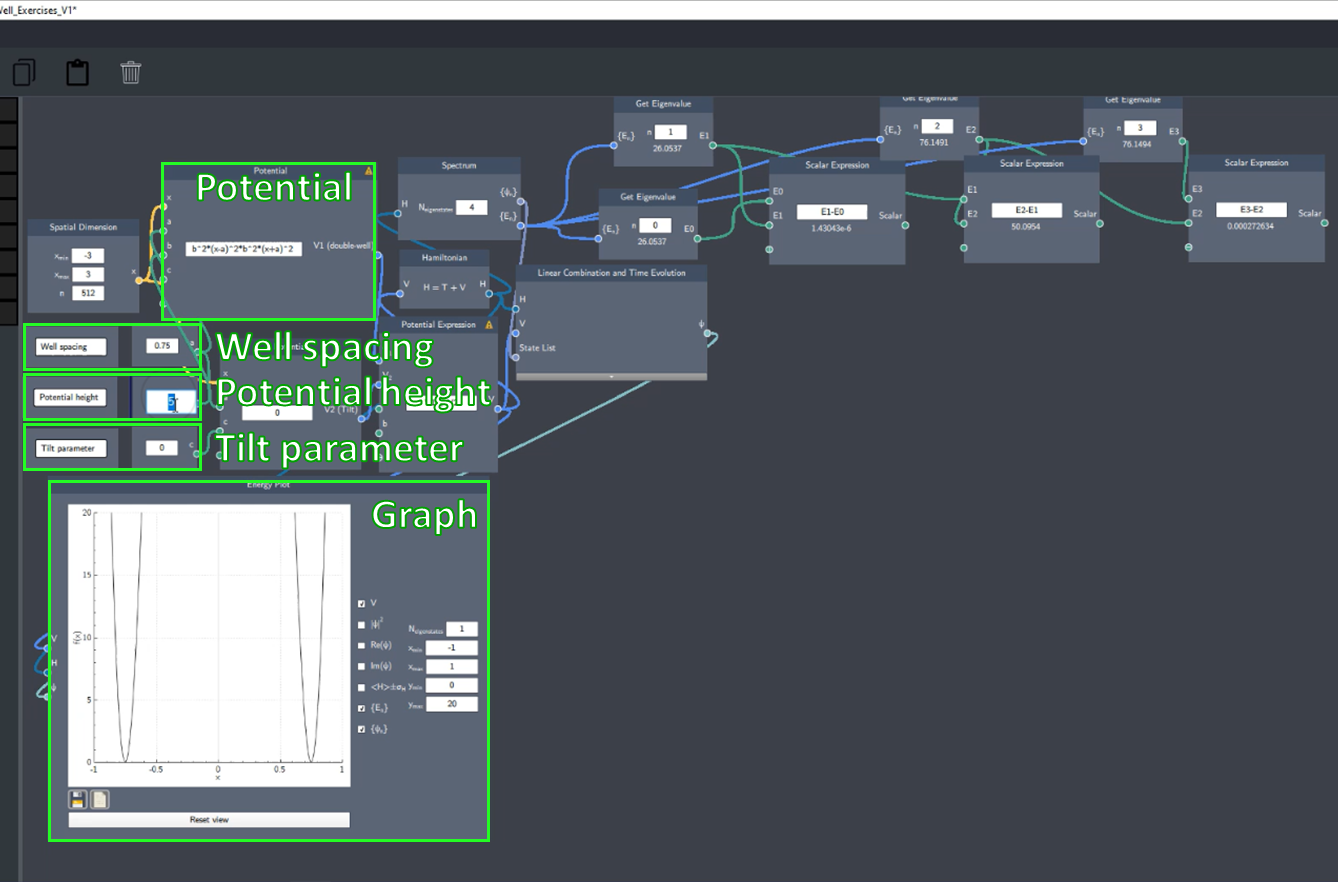}
\caption{The Composer \emph{flowscene} describing the double-well system used in this study. In green, we highlight the five areas of interest (AOIs) analyzed for this study. Starting from the top and moving down, we highlight the \emph{Potential node} that defines the double-well potential via the expression entered in the box (Eq.~\eqref{Eq:Pot}). We also highlight the parameters that the participant can manipulate by changing the numbers in the \emph{Scalar Expression} nodes defining the double-well spacing, the height of the potential, and the tilt parameter defining the relative offset of the two wells. Finally, we highlight the \emph{Energy Plot} node that visualizes the potential, the energy levels, and the energy eigenstates. In the top right-hand side of the \emph{flowscene}, we show the energy difference $E_{k+1} - E_k$ for energy eigenstates $k = 0, 1,$ and $2$. The \emph{flowscene} is arranged in the same way for all participants in order to reliably compare eye-tracking data between participants.}
\label{fig:AOIs}        
\end{figure*}

In this subsection, we describe Quantum Composer and situate it within the existing landscape of educational tools for visualizing and simulating quantum systems. Quantum Composer (or simply Composer) was developed at Aarhus University, and it provides a graphical interface through which one can access the computational power of the QEngine, a C++ library for the simulation and control of quantum systems~\cite{sorensen2019qengine}. Composer, described in detail in Ref.~\cite{quatomic}, can be downloaded for free for most desktop platforms~\cite{quatomic}. It is designed as a flow-based tool where users can drag-and-drop different \emph{nodes} that define and visualize static and dynamic quantum mechanical systems. In what follows, for consistency, we maintain the Composer terminology introduced in Ref.~\cite{zaman2021quantum}. These collections of \emph{nodes} and their interconnections defines a \emph{flowscene} that simulates and visualizes the system of interest. For example, Fig.~\ref{fig:AOIs} shows the \emph{flowscene} used in this study where a static, double-well potential is explored.

A number of educational simulation tools in quantum mechanics exist, each with varying degrees of validation. It is Quantum Composer's flexibility that makes it different from other such tools. Therefore, it represents a bridge between more specialized libraries written for programming languages (e.g., Python's QuTIP~\cite{johansson2012qutip,babayev2021computational}) and specific graphically oriented applets (e.g., those found in Refs.~\cite{Falstad, Quantum_online, christian2015physlet}). When taken in conjunction with the exercises provided on the Composer website~\cite{quatomic}, which come with pre-made \emph{flowscenes}, Composer is very similar to the QuILT~\cite{sayer2017quantum} and QuVIS~\cite{kohnle2018quvis} projects in that these exercises revolve around pre-defined systems where students are expected to make minimal changes to the simulation interface. Notable also are the quantum-mechanics-based PhET simulations~\cite{mckagan2008developing}; PhET simulation design is greatly influenced by research done on their interfaces~(PhET.general.2, PhET.general.3, PhET.general.4). Additionally, the QuILTs have been validated across a range of topics in quantum mechanics~\cite{zhu2011improving,sayer2017quantum,justice2019improving}. Similar research has been carried out using Composer~\cite{Carrie_2020}, suggesting that students do benefit from the visualizations found within the tool. The study described here is intended to determine the differences between how experts and novices use Quantum Composer, with the goal of both demonstrating how the information given by Composer is used and developing the tool to better facilitate teaching and learning in quantum mechanics. Quantum Composer is particularly useful for the gaze study presented here due to the programmability of the Composer flowscenes. This means that the scientific problems optimal for the study can be programmed directly rather than having to select from the pre-existing simulations. In addition, as described below, the visual composition of the interface has dramatic consequences for the information processing during task solution. Again, the programmability of the composer interface allowed the research team to generate the interface structure hypothesized to be optimal for the concrete study as well as - in the future- allowing for experimentation of modifications of the interface.

\subsection{The role of multiple representations for teaching}
For scientific learning, multiple external representations (MERs) play a beneficial role, which is well documented for the natural sciences \cite{tytler2013constructing} and physics \cite{treagust2017multiple}. MERs are especially important for conceptual understanding \cite{dooren2010addition} and are discussed as a necessary condition for in-depth understanding \cite{disessa2004coherence}. Ainsworth created a conceptual framework that provides an overview of the prerequisites for the effective use of MERs in teaching and learning situations and the unique benefits for learning complex or new scientific content \cite{ainsworth2006deft}. According to Ainsworth, in the design, functions, and tasks (DeFT) taxonomy, learning with MERs means that two or more external representations (e.g., diagrams, formulas, and data tables) are used simultaneously. Specifically, there are three key functions that MERs can fulfill (even simultaneously) to support the learning process. According to the first function, MERs can complement one another either by providing complementary information or by allowing for complementary approaches to process information. The presentation of MERs with complementary information content may be advantageous if the presentation of all relevant information in a single form of representation would lead to cognitive overload of the learner. However, even if different representations contain the same information, they can support the learning process by allowing learners to select the most appropriate representation to accomplish the given learning task in the particular learning situation. The second function is that simultaneously presented representations can constrain one another's interpretation in two ways: A more familiar representation can constrain the interpretation of a less-familiar one, or the inherent properties of one representation can trigger the usage of another representation, which is considered helpful for the learning process. According to the third function, the construction of a deeper understanding is fostered if learners integrate information from different forms of representation to gain insights that could not have been obtained with just one form of representation. Although MERs demonstrably have the potential to support learning processes \cite{ainsworth2006deft}, their use in learning situations is also associated with learner demands that can increase learners' cognitive load and even negatively affect the learning process \cite{swaak1998supporting}. Indeed, many studies point toward student difficulties with MERs (e.g., \cite{ainsworth2006deft}; \cite{nieminen2010force}. Consequently, the cognitive load of the learning environment must be considered and managed carefully. In this context, technology support can help reduce cognitive load when learning with MERs, and it can therefore facilitate the learning-promoting effect of MERs \cite{horz2009differential}.

When students extract information from MERs to reason with them, they specifically need metarepresentational competence. DiSessa defines representational competence as the ability to create representations, critique and compare the adequacy of representations, understand the purposes of representations and explain representations \cite{disessa2004coherence}. Therefore, this competence is very relevant for solving reasoning tasks with Quantum Composer. In previous work at the intersection of mathematics and quantum mechanics, Wawro, Watson and Christensen categorized students metarepresentational competence according to value-based preferences, problem-based preferences and purpose and utility awareness and used it as a measure for students' ability and flexibility in using different algebraic notations \cite{wawro2020students}. The authors demonstrated that student understanding of linear algebra is supported by the flexible use of different notational systems, and the authors suggest that every instruction in quantum mechanics should provide opportunities for students to improve their metarepresentational competence. 

\subsection{Gaze behavior of experts and novices}
The underlying assumption when analyzing learners' eye movements is the Eye-Mind hypothesis \cite{just1976eye}, which states that learners process the information they are focusing on. Following this assumption, it was found that students who are unfamiliar with representations and who have a low content knowledge (novices) process information differently than people with high content knowledge (experts) \cite{larkin1980expert}. This difference can be explained by a number of reasons. On the cognitive level, it was found that experts process information in chunks, which means that they process elements that belong together as a single element, and therefore, they are more efficient at information processing \cite{simon1974mirrors}. Similarly, experts are also more efficient at selecting relevant and neglecting irrelevant information than novice learners. This so-called information reduction hypothesis was originally stated by Haider and Frensch (1999), and it was confirmed in a meta-analysis by Gegenfurtner and colleagues \cite{gegenfurtner2011expertise}. In terms of gaze behavior, this means that experts require less time to identify relevant information and have a shorter total fixation duration on relevant information. Another aspect of the superior information-processing behavior by experts is explained by the theory of long-term working memory (Ericsson and Kintsch, 1995), which states that experts possess retrieval structures which allow a faster recall of information stored in long-term working memory. On the perceptual level, the holistic model of image perceptions states that experts possess superior parafoveal processing (i.e., processing of visual information outside the foveal region on the retina), which allows them to relate information that is spatially further separated because they can use the parafoveal regions in the eye for information extraction. In terms of eye movements, this means that experts would have longest saccades, which are fast eye-movement between two fixations.

For education, the analysis of the gaze behavior of experts and novices has received a lot of attention \cite{larkin1980expert} because the gaze behavior of experts may be used to design instructional material in case there is a relation between the perceptual processing and the conceptual understanding or learning-relevant competences. 
Graphs are one type of representation in which the perceptual processing is closely related to the conceptual understanding. This conclusion is evident from the finding that various machine learning algorithms are able to predict student performance based on gaze behavior on the graph surface during problem solving \cite{kuchemann2020classification}, and during conceptual learning \cite{dzsotjan2021predictive} in physics. In a recent review on eye tracking, K\"uchemann and colleagues identified 27 publications between 2003 and 2020 that focus on students' gaze behavior on graphs \cite{kuchemann2020gaze}. The authors found that the efficiency in extracting information from a graph is not caused by the content knowledge but by the familiarity to the graph, and a lower content knowledge was related to more transitions between answer options in the graph area (K\"uchemann et al., 2022).

In the context of expert - novice comparison, Okan et al. also verified the information-reduction hypothesis in graphs (Okan et al., 2016). The authors observed that experts are more efficient in identifying task-relevant areas in a graph, which allows them to invest more time in evaluating the relevant information. Specifically, experts verified axis labels and scaling more often to avoid mistakes. Furthermore, Andr\'a and colleagues analyzed the gaze behavior of university students with different levels of previous knowledge in mathematics during problem-solving with graphs, texts, and equations \cite{andra2015reading}. The authors found that students have the highest mean fixation duration on the equation among these three types of representations, which suggests that equations are most difficult to process in comparison to the two other representations. In physics education research on a single representation, it was found that students who determine the slope of a graph correctly focus longer along the graph \cite{susac2018student, klein2019student, bruckner2020epistemic} and perform more transitions between the trend of the graph and the axis values \cite{kuchemann2020students}. 

As mentioned above, simulation tools often display multiple external representations simultaneously, which need to be understood, integrated, and brought into the context of the task. In this context, it is an open question to what extent previous knowledge, visual behavior, representational competences influence students' performance while solving tasks with a simulation tool. 

\section{Methods}
\subsection{Participants}
In total, 36 participants (29 male, 6 female, 1 no statement) from the TU Kaiserslautern, the WWU M\"unster, and the Munich Center of Quantum Science and Technology (MCQST) took part in the study. Among the participants, there were 28 master students in physics (21 male, 6 female, 1 no statement) who had completed at least one course in quantum mechanics, and 8 PhD students or postdocs with a degree in physics. For data analysis, the students were assigned to the group of novices or experts depending on the relative pretest score $p$ (experts $p>0.5$ , see below for details). This resulted in the final number of 28 novices and 8 experts. The latter group consisted of 5 PhD students and postdocs as well as 3 master students, whereas the group of novices consisted of 25 master students and 3 PhD students and postdocs. The study was conducted in German, and all English terms in Quantum Composer were translated on a sheet of paper handed out to the participants. Participation was voluntary and compensated with 15 EUR at TUK and MCQST, the students at WWU M\"unster received no compensation, and B2 was required as a minimum language level in German. 

\subsection{Experimental design}
At the beginning of the session, an instructor explained the course of the study and the details of the data recording to the participants. Then, the students gave their informed consent to the data recording as well as the privacy and data protection regulations. Afterwards, the students watched an instructional video of the theoretical background of the harmonic oscillator before completing the pretest. Then the students started working with Quantum Composer. Here, the students first solved an example task that was unrelated to the double-well potential or the tunneling effect to get familiar with Quantum Composer. Before this example task, all students had not seen Composer at all. They had five minutes to do this. Afterwards, the eye trackers were calibrated, and the students had a maximum of 20 minutes to complete six Quantum Composer tasks. The tasks targeted phenomena in the double well potential where the participants had to adjust the wells by using the parameters like \emph{well spacing} or \emph{potential height} as well as the parameters of the automatically generated graphs.

\subsection{Pretesting and Quantum Composer tasks}

\subsubsection{Pretest}
Though a pre-screening of the participants regarding their backgrounds in quantum physics was done, a pretest was used to assess whether the participants were experts or novices regarding the double-well potential and its linked concepts. For this, five questions were chosen from existing instruments measuring basic ideas of quantum physics like the potential well or the interpretation of wavefunctions \cite{mckagan2010design,sadaghiani2015quantum,zhu2012surveying}. As the questions were single-choice, in the end the score obtainable in this part was five points. In addition, a drawing task was devised where the participants had to show what knowledge they already had in the context of the double-well potential and its energy levels and wave functions. This drawing task was evaluated by determining whether certain aspects were drawn correctly (netting a point) or drawn incorrectly/ not at all (netting zero points). The drawings of the first two eigenstates as well as the first two eigenenergies for two different cases of the well were coded, adding up to eight points obtainable in the drawing task. Additionally, the participants had to answer an additional open-ended question about the relationship between potential height and eigenenergies with two obtainable points for relating the two correctly.  \\
In summary, there were 15 points obtainable, five of them being from very basic questions about quantum physics and 10 being from questions regarding the problem used as an example in this study - the double well. While the first part was an adaption of previously established instruments, the questions regarding the double-well were newly developed. All new questions were piloted by a cohort of students regarding their clarity. The answers obtained in the survey were then coded by two independent experts using the same coding manual and the reliability of the codings was evaluated by using Cohen's Kappa and the recommendation by \cite{landis1977measurement}. The values obtained are all very high and almost perfect (see Tab. \ref{tab:coh}), which was to be expected by the low inferent nature of the category system used. All in all, 43 answers were categorized for each aspect. To be deemed an expert, at least 50\% of the points had to be obtained (i.e., 7.5 points). \\

\begin{table}[]
\caption{Values of Cohen's Kappa calculated after coding of all answers to the double-well problem by two independent experts.}
\begin{tabular}{lc}
\hline \hline
Category                        &~~~ Cohen's Kappa \\ \hline
\vspace{-2mm}  &  \\
Eigenenergy ground state    & 1.00          \\
Eigenenergy first excited state & 1.00          \\
Eigenstate ground state          & 0.95          \\
Eigenstate first excited state  & 0.98         \\\hline \hline
\end{tabular}
\label{tab:coh}
\end{table}

\subsection{Quantum Composer tasks}
In Quantum Composer tasks, the participants' basic concepts of quantum physics in the context of the double-well potential with the following potential function were tested:
\begin{equation}
\label{Eq:Pot}
    V(x)=b^4 \cdot (x-a)^2 \cdot (x+a)^2 + c\cdot x. 
\end{equation}
In this equation, $2\cdot a$ corresponds to the well spacing, $b$ is related  to the height of the potential barrier (due to $V(0)=b^4a^4$), and $c$ is a tilt parameter, that was equal to zero in the first two tasks. The tasks were composed in the form of slides containing instructions and questions that had to be performed sequentially. Participants were able to switch between the current task slide and the Quantum Composer application which they could use to perform the experiment according to the task. The answers were given verbally after students communicated their readiness, and these were audio recorded. After participants completed their answers, they could move on to the next task. In total, they had 20 minutes to complete up to six tasks. If the 20 minutes ran out while the answer was being given, participants were allowed to finish the explanation. The first task was completed by all 36 participants, 34 of them finished the second task, 14 answered the third task, 11 completed the fourth task, three were able to respond to the fifth task, and only one participant even gave an answer to the sixth task. 

 The concepts addressed were wavefunctions, eigenstates, eigenenergies, symmetries and tunneling. In this work, we focus our statistical analysis on tasks 1 and 2, because they were completed by most of the participants.
 The Quantum Composer tasks were:
 \begin{enumerate} 
 \item Consider the two lowest lying energy states of a system in a double-well potential. Is the ground state symmetric or anti-symmetric?
 \item How does the splitting of the ground state and first excited state of the system qualitatively depend on changing the trap potential parameters, i.e., potential height and well spacing?  
 \end{enumerate}
At the end, the students answered demographic information (gender, degree program, university) and received their compensation, if applicable.  

\subsubsection{Analysis of answers to Quantum Composer tasks }
To evaluate the understanding during the Quantum Composer tasks, the answers to each task were recorded whilst the Composer software was made unavailable to discriminate between the task-solving and the response process. 
The answers were transcribed and coded for their correctness by using a deducted category system. The category system was tested for reliability by using a coding manual applied by an independent interrater on 8 of the 43 answers that were randomly selected. The reliability was calculated by using Cohen's Kappa ($\kappa=0.85$). Thus, the coding was deemed reliable. All in all, 12 aspects had to be addressed during the verbal responses (seen in Tab. \ref{tab:cat}). For a wrong statement regarding one of the aspects, $-1$ points were given, for not answering, zero points were given and for correctly answering, one point was given. Thus, the amount of points achievable in the Quantum Composer tasks ranged from -12 to 12.

\begin{table}[]
\caption{Categories coded for each of the six tasks given to the participants during the work with Composer.}
\begin{tabular}{cl}
\hline \hline
Task~~~~ & ~~Aspect addressed       \\ \hline
\vspace{-2mm}  &  \\
1             & ~~Potential height               \\
1             & ~~Well-spacing                   \\
1             & ~~Explanation                    \\
2             & ~~Symmetry of the wavefunction  \\
2             & ~~Change by changing parameters  \\
2             & ~~Explanation                    \\
3             & ~~Dependency on tunnel-frequency \\
3             & ~~Explanation                    \\
4             & ~~How to suppress oscillations   \\
4             & ~~Explanation                    \\
5             & ~~Tilt-parameter chosen          \\
6             & ~~Periodicity                   \\ \hline \hline
\end{tabular}
\label{tab:cat}
\end{table}

\subsection{Eye Tracking}
To record participants' eye movements, we used a Tobii Pro Fusion eye tracker with a frequency of 120 Hz, a nominal spatial accuracy of 0.3$^\circ$, and nominal precision of 0.2$^\circ$). The software Tobii Pro Lab was used to control the eye tracker. The stimuli were presented on a screen with 1920$\times$1080 pixels and a refresh rate of 75 Hz. Prior to the start of the study, we performed a nine-point calibration and introduced the participants to the basics of eye tracking. 

During Quantum Composer tasks, the eye movements were recorded on the entire screen and not on specific stimuli. For the data analysis, we used the assisted mapping tool in Tobii Pro Lab, supported by manual mapping to assign the gaze points to a screenshot of Quantum Composer (see Fig. \ref{fig:AOIs}).

For the data analysis, we analyzed five areas of interest (AOIs; see Fig. \ref{fig:AOIs}): the potential, the well spacing, the potential height, the tilt parameter, and the graph. The potential contains information about the underlying physical potential in which the wave function is displayed. To remind the reader, the well spacing ($2a$ in Eq. \ref{Eq:Pot}) is a parameter in the potential that changes the distance between the two wells. The potential height ($b$ in Eq. \ref{Eq:Pot}) is a parameter in the potential that controls the height of the potential barrier between the two wells. The tilt parameter ($c$ in Eq. \ref{Eq:Pot}) controls an asymmetry in the double well potential. We focus on these five AOIs because they were relevant for the first two tasks of Quantum Composer which were completed by all participants whereas the other areas almost received no visual attention even though Task 1 could be solved with the numeric values of the energy states.
For the statistical comparison between experts and novices, we used the non-parametric Wilcoxon rank sum test.

\section{Results}
\subsection{Performance}
\begin{figure}[t!]
\centering
\includegraphics[width=\linewidth]{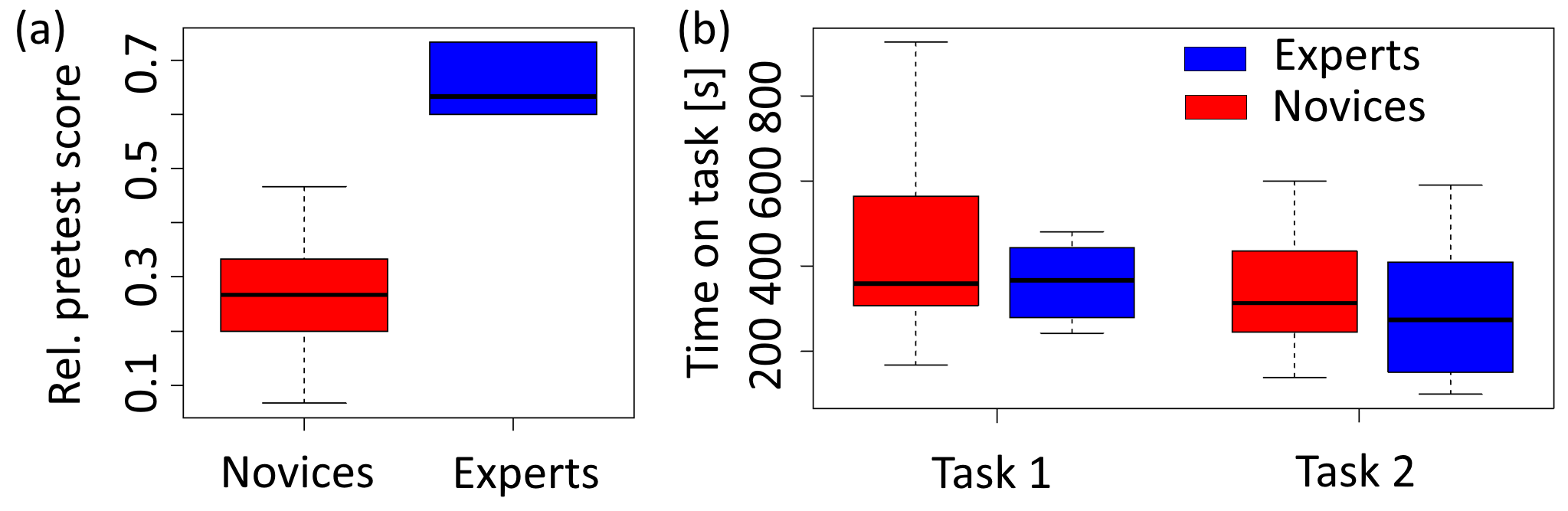}
\caption{Box plots of (a) the relative pretest score of experts and novices, and (b) the time on the Quantum Composer tasks of experts and novices in Task 1 and Task 2. The black lines in the boxes indicate the median, the boxes have the sizes of the lower and upper quartile, and the whiskers are located at the last data point within 1.5 times the interquartile range, i.e., the distance between the upper and lower quartile.}
\label{fig:Pre_ToT}    
\end{figure}

After assigning the participants into the expert and novice groups based on the pretest score as mentioned above, the novices (experts) had an average score of 0.28 (0.63) in the pretest (Fig. \ref{fig:Pre_ToT}a).
The novices needed 444 $\pm$ 187 s, and the experts needed 363 $\pm$ 95 s ($p>0.05$, Fig. \ref{fig:Pre_ToT}b) to complete Quantum Composer Task 1. For Task 2, the novices required 346 $\pm$ 128 s, and the experts needed 294 $\pm$ 174 s ($p>0.05$).

\begin{figure}[t!]
\centering
\includegraphics[width=\linewidth]{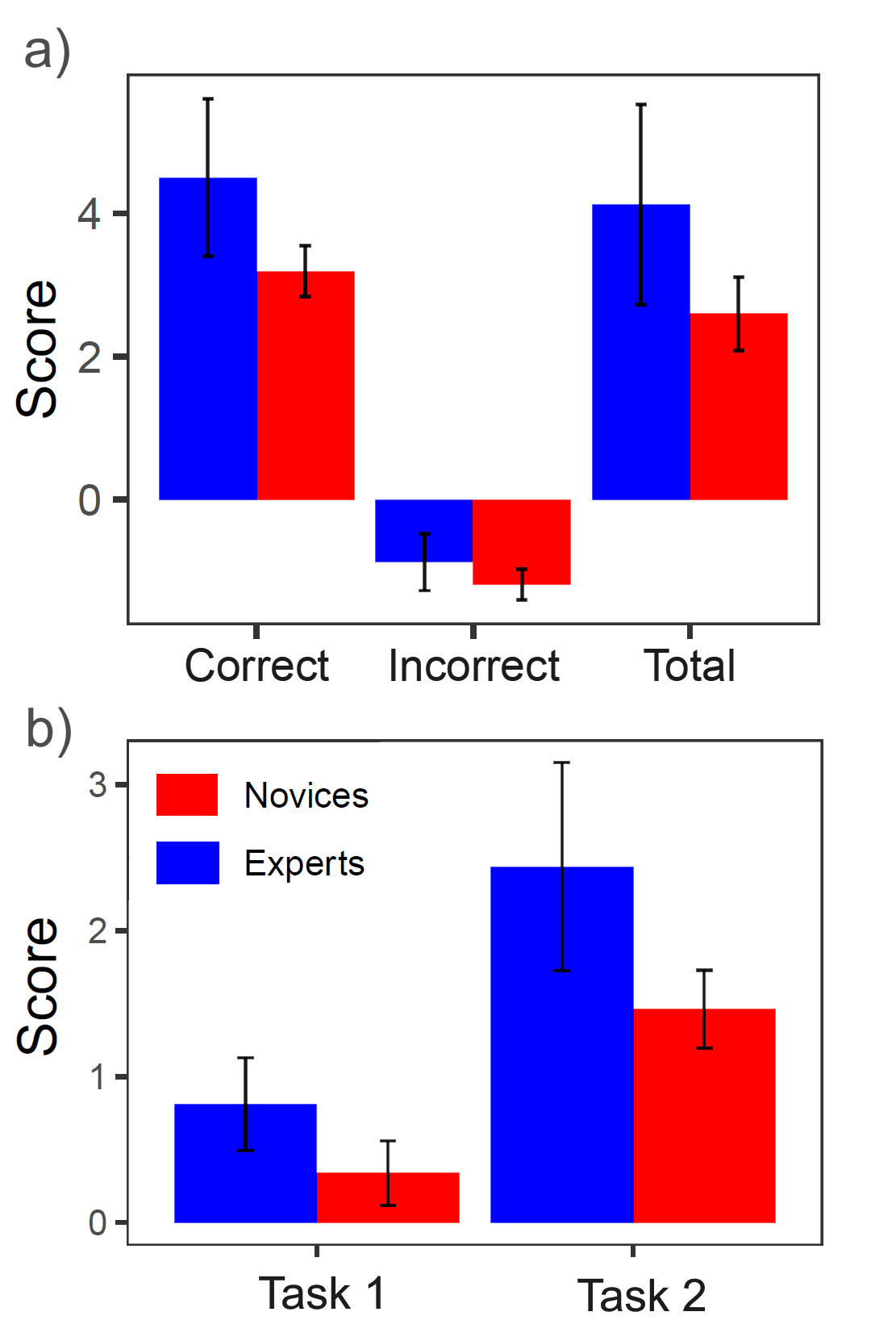}
\caption{(a) The correct, incorrect, and total scores during Quantum Composer tasks of experts and novices. The participants could reach a maximum score of 12 points (correct) and a minimum score of -12 points (incorrect). (b) Scores of experts and novices in Task 1 and Task 2. In both tasks, the score ranged from -3 to 3 points.}
\label{fig:Scores}    
\end{figure}

To determine the participant scores during the Quantum Composer tasks, we rated the answers of the participants regarding the correct statements (positive scores) and the incorrect statements (negative scores) which we used to determine the total score (Fig. \ref{fig:Scores}a). We found that experts had more correct answers and fewer incorrect answers, thus giving them a higher total score. We also found that experts reach a higher score in Task 1 ($p>0.05$) and in Task 2 ($p=0.040$, Cohen's $d=0.76$, Fig. \ref{fig:Scores}b). In each task, the scores ranged from -3 to 3 points. However, all differences are not statistically significant.

Since we did not find a significant difference in the total score of experts and novices, we studied a potential mediation effect of the ability to reason with the visuals representations in Quantum Composer, i.e., participants' metarepresentational competences.

To quantify participants' metarepresentational competence and participants' reasoning with Quantum Composer, we analyzed how participants related their conceptual knowledge to the information from Quantum Composer by counting the number of concepts the participants mentioned during their answers. Based on the transcripts of the answers of the participants, we deduced ten different concepts used during the argumentation. We found that on average four different concepts were mentioned per interview. The concepts mentioned during the answers and corresponding occurrence rates are presented in Tab. \ref{tab:concepts}:

\begin{table}[]
\caption{Concepts mentioned during the answers to the Quantum Composer tasks}
\begin{tabular}{lc}
\hline \hline
Concept                       & Occurence rate \\ \hline 
\vspace{-2mm}  &  \\
 Amplitude / minima /     & $83 \%$         \\
\vspace{2mm} maxima of wave functions &  \\ 
\vspace{2mm} Ground state/ excited state & $66\%$ \\
\vspace{2mm}Eigenmodes / eigenstates & $55\%$ \\
\vspace{2mm}Comparison with harmonic oscillator & $42\%$\\
\vspace{2mm}Nodes / anti-nodes  & $42\%$ \\
\vspace{2mm}Symmetry considerations & $33\%$ \\
\vspace{2mm}Linear / nonlinear behavior & $22\%$ \\
Probability considerations / & $22\%$ \\
\vspace{2mm} evanescence in a potential & \\
\vspace{2mm}Superposition principle & $19\%$ \\
Wavelength & $5\%$ \\ \hline \hline
\end{tabular}
\label{tab:concepts}
\end{table}


We found over all tasks, that the experts referred to a significantly higher number of concepts to explain the answers ($p=0.007$,  $d=1.10$).

\begin{figure}[h!]
\centering
\includegraphics[width=\linewidth]{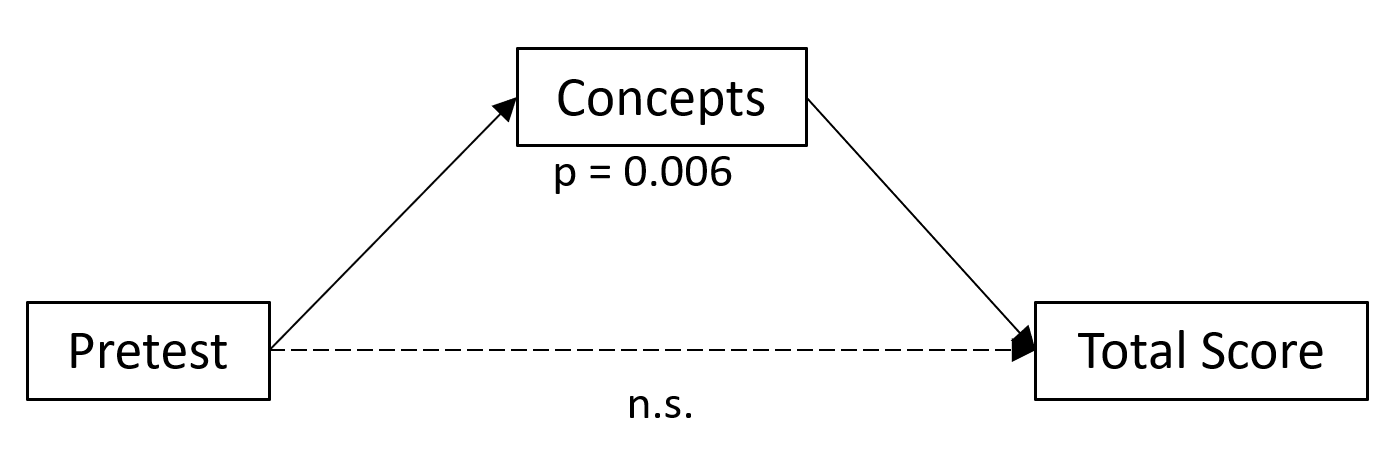}
\caption{Diagram of the relation between the pretest and the total score. There is no significant direct effect of the pretest score on the total score, but a significant indirect effect with the number of concepts mentioned during the tasks as a mediator.}
\label{fig:Mediation}        
\end{figure}

To understand the relationship between the pretest, the number of concepts used, and the total score, we used an ordinary least squares (OLS) regression-based approach for conditional process modelling by Hayes \cite{hayes2017introduction}. We used a Quasi-Bayesian Confidence Intervals approximation and performed 1000 Monte Carlo simulation draws during the mediation analysis. 
Within the mediation analysis with the pretest as the independent variable, the number of concepts as the mediator, and the total score as the dependent variable, we found a significant average causal mediation effect of 4.0 ($p=0.006$, proportion mediated: 0.76, $p=0.05$) and an insignificant direct effect ($p=0.64$). This relationship is shown in Fig. \ref{fig:Mediation}. 

\subsection{Parameter modification}
To assess whether experts and novices use Quantum Composer differently to solve the tasks, we counted the times participants modified the Quantum Composer parameters (i.e., the well spacing, potential height, and the tilt parameter). In Task 1 and 2, we found no difference in the number of modifications in the parameters defining the well spacing, the potential height, and the tilt parameter ($p>0.05$). In both tasks, we found a tendency for novices to perform a higher number of modifications of the scaling of the graph ($p>0.05$).

\begin{figure}[h!]
\centering
\includegraphics[width=\linewidth]{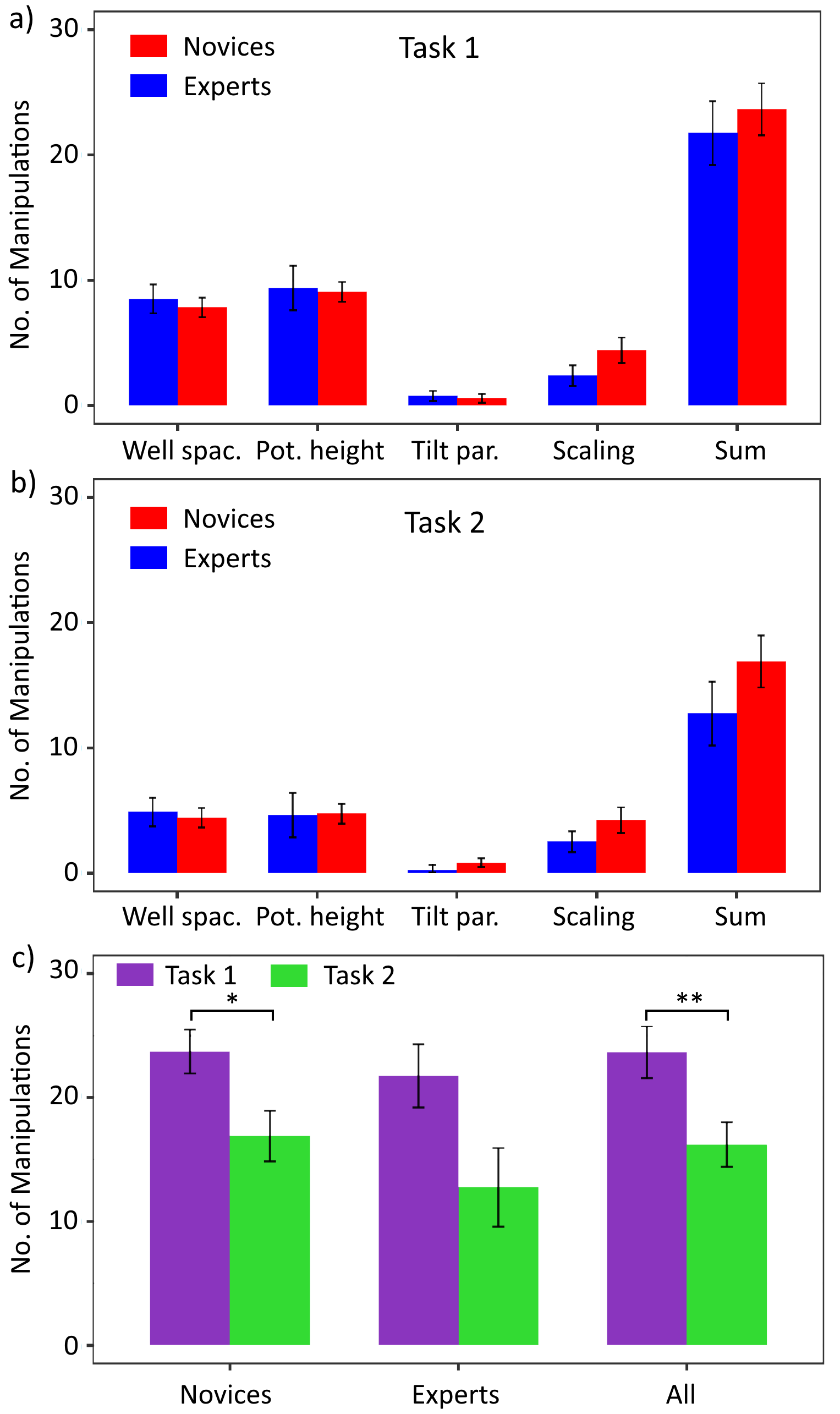}
\caption{Number of manipulations of the well spacing, the potential height, the tilt parameter, the axes scaling, and the total number of manipulations in Task 1 (panel a), Task 2 (panel b), as well as a comparison between these two tasks (panel c). The label $*$ indicates a significant difference with $p<0.05$ and $**$ means $p<0.01$.}
\label{fig:No_Man}        
\end{figure}

\subsection{Gaze behavior during Quantum Composer Tasks}
\begin{figure}[h!]
\centering
\includegraphics[width=\linewidth]{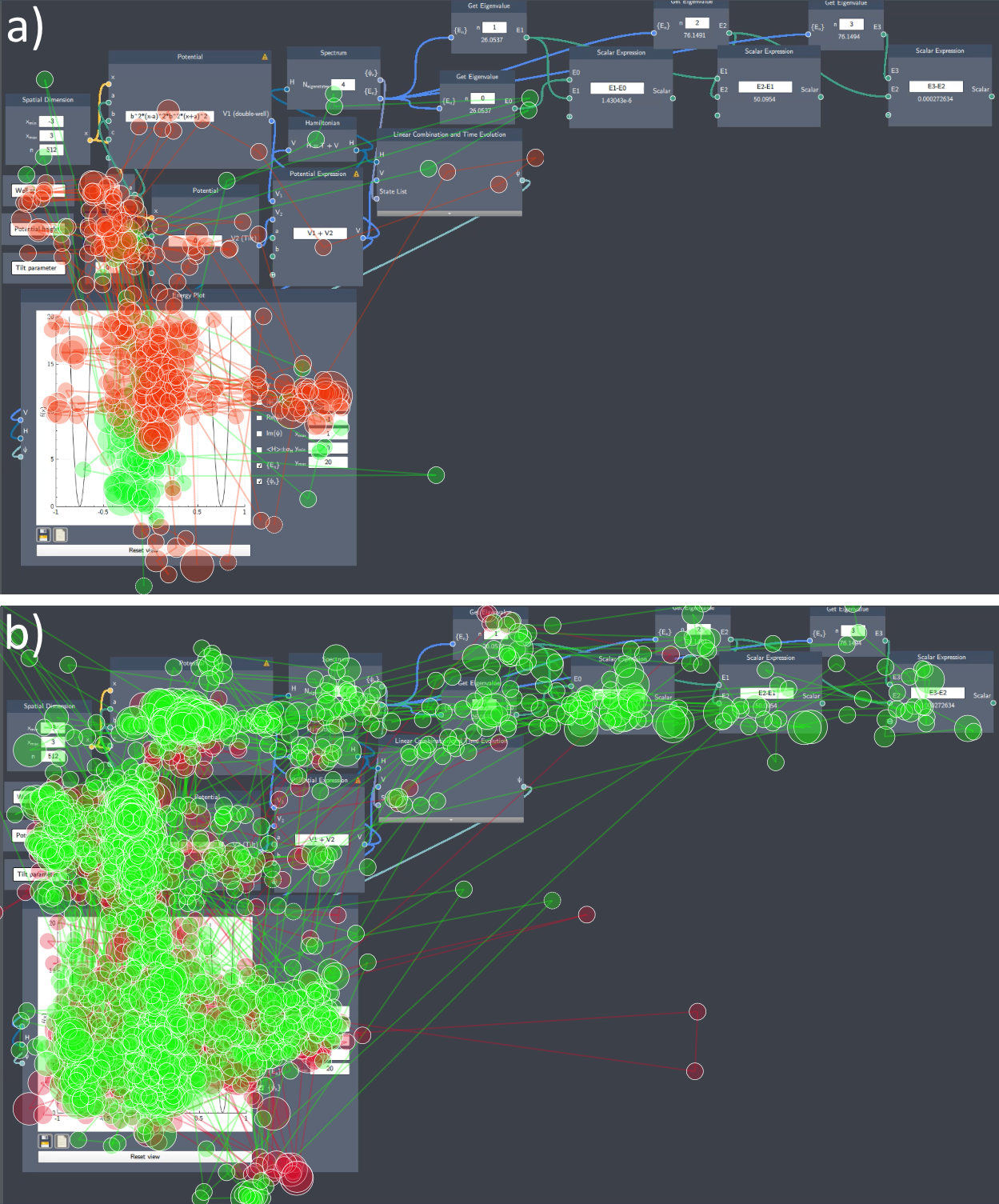}
\caption{Gaze patterns in red and green in Task 2 of the two experts with the highest total scores (a) and two novices (also in red and green) with the lowest total scores (b). The circles indicate fixations and the sizes of them visualize the fixation durations. The lines between two fixations are saccades, i.e., rapid eye movements between fixations.}
\label{fig:GPs}        
\end{figure}

To obtain a qualitative impression of the task-solving behavior of experts and novices, we compared the gaze patterns of the two best-performing experts (each of them achieved a total score of 11 out of 13 points) and the two lowest-scoring novices (total score of 1 and 2 out of 13 points) while solving Task 2 (Fig. \ref{fig:GPs}). 
It is noticeable that the former have a clearly lower number on fixations on all areas. The selected novices examine all areas, relevant and irrelevant, whereas the experts only focus on the relevant areas.

We analyzed participants' relative dwell time (normalized by the total dwell time on all parts during one task) on specific relevant and irrelevant parts of Quantum Composer interface. For solving Tasks 1 and 2, it is necessary to manipulate the well spacing, the potential height, but not the tilt parameter or the potential function. The effects of these manipulations are visible either in the graph, where the ground and excited states as well as the potential and the wave functions are visualized. Alternatively, it would be possible to extract the energy splitting of the ground and first excited state from the numerical values. However, the numerical values of the energy difference between the two states received only little attention by any participant, so that a statistical comparison was not possible.

\begin{figure}[h!]
\centering
\includegraphics[width=\linewidth]{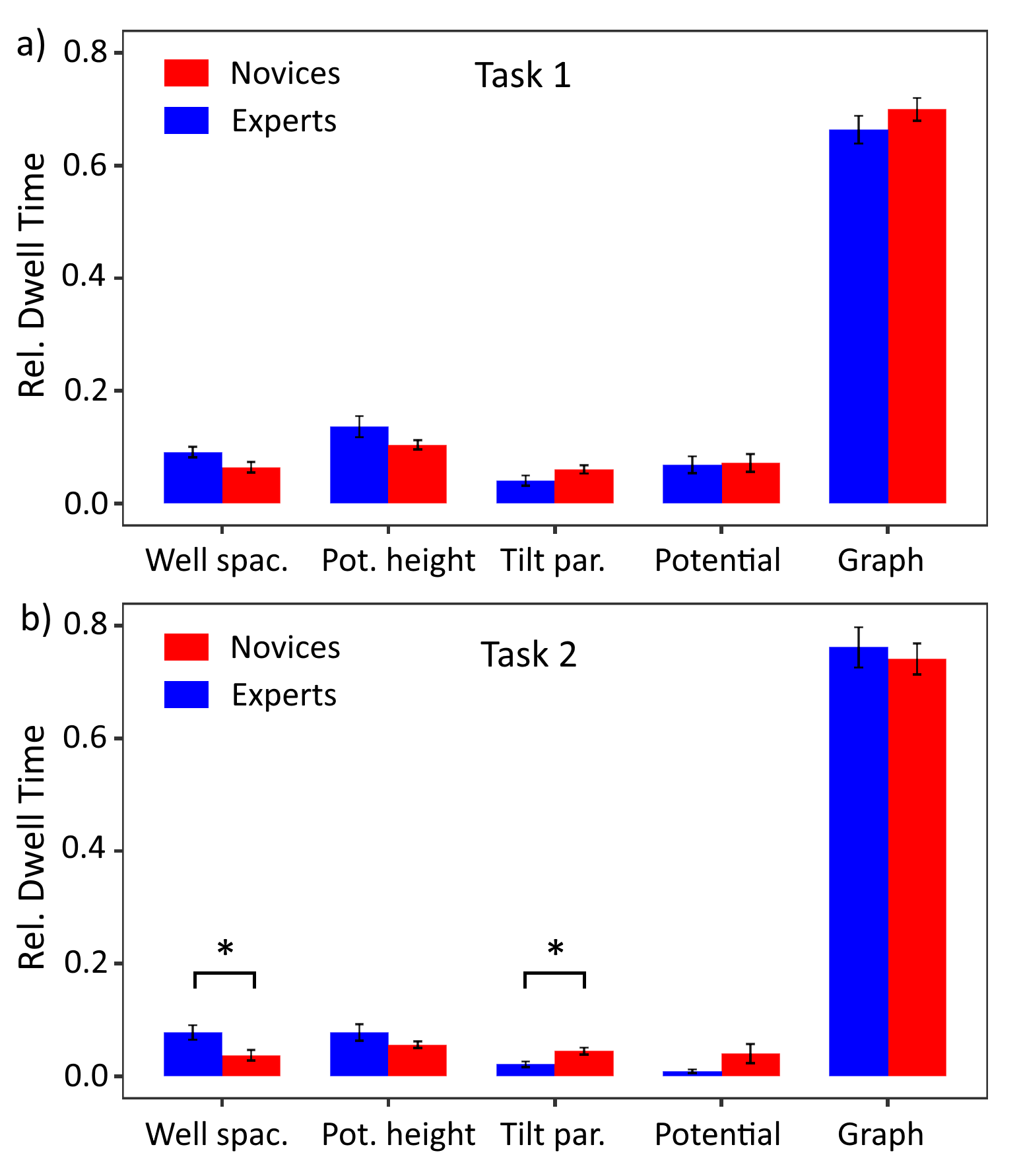}
\caption{Participants' relative dwell time on five relevant and irrelevant areas in Task 1 (a) and Task 2 (b): the well spacing, the potential height, the tilt parameter, the potential function, and the graph. The $*$ indicates a significant difference with $p<0.05$}
\label{fig:ET}        
\end{figure}

Fig. \ref{fig:ET} shows the relative dwell time of novices and experts in Task 1 (panel a) and Task 2 (panel b). It is noticeable that the graph in both tasks received the most attention by novices (Task 1: 70.1\%, Task 2: 74.4\%) and by experts (Task 1: 66.3\%, Task 2: 76.1\%) with a non-significant difference between the two groups. Apart from that, in Task 1, there is a tendency for higher relative dwell time by experts on the relevant parameters such as the well spacing ($p>0.05$) and the potential height ($p>0.05$), and a non-significant higher relative dwell time on the tilt parameter by novices ($p>0.05$). In Task 2, there is a significantly higher relative dwell time on the well spacing ($p= 0.013$, $d=0.87$) and a tendency towards a higher dwell time on the potential height ($p>0.05$) for experts. Apart from this, there is a significantly lower dwell time on the tilt parameter ($p= 0.018$, $d=0.80$) and a tendency towards a lower dwell time on the potential function ($p>0.05$) for novices. 

\begin{figure}[h!]
\centering
\includegraphics[width=\linewidth]{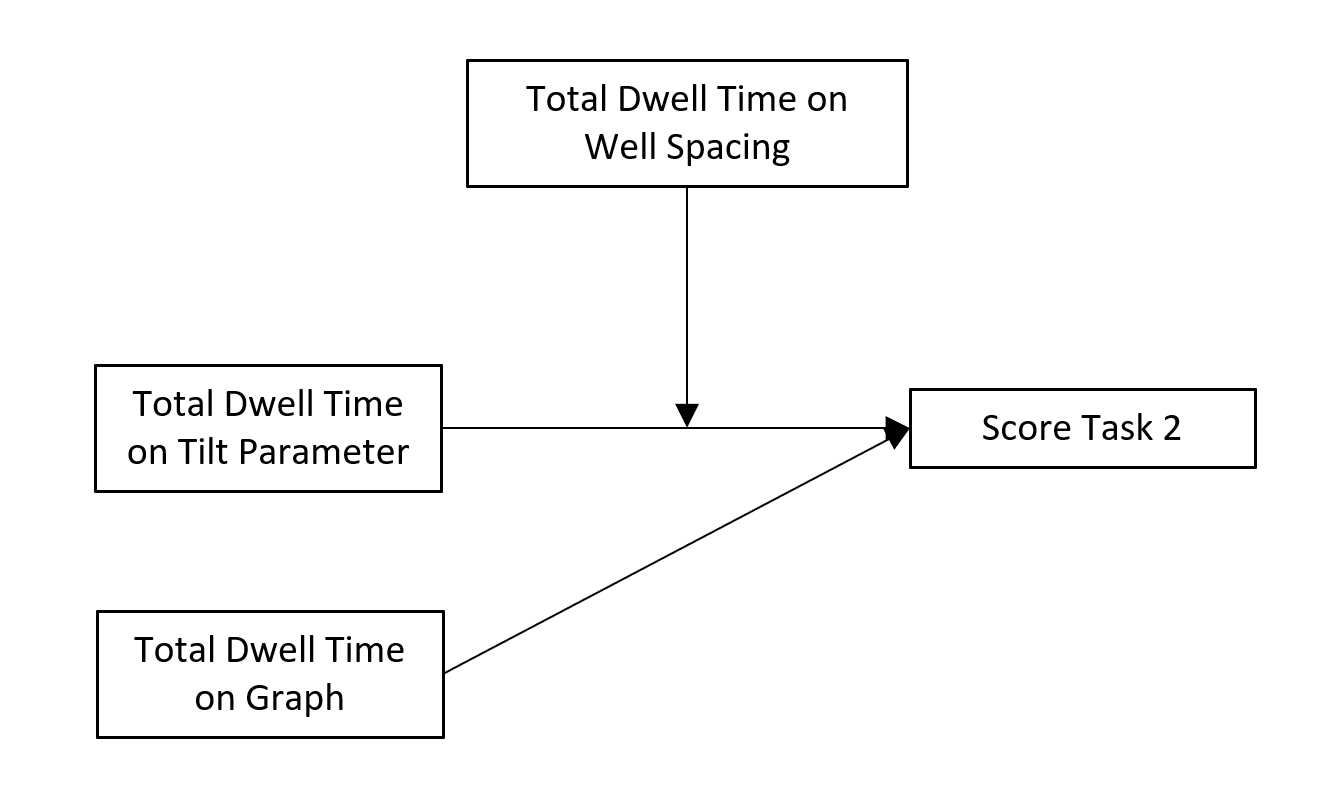}
\caption{The total dwell time on the graph and the total dwell time on the tilt parameter both exhibit a linear relationship with the score in Task 2, but the effect of the total dwell time on the tilt parameter is moderated by the total dwell time on the well spacing.}
\label{fig:Moderation}        
\end{figure}

To relate these results in Task 2, we performed an OLS multiple linear regression and found that there is a significant linear relationship between the total dwell time on the graph ($p=0.001$) and the total dwell time on the tilt parameter ($p=0.003$) as antecedent variables and the score in Task 2 as the dependent variable. In this linear model, there is a significant moderation effect of the total dwell time on the well spacing on the relation between the total dwell time on the tilt parameter and the score in Task 2 ($p=0.037$) (Fig. \ref{fig:Moderation}). This means that for participants who focus longer on the well spacing, the relationship between the dwell time on the tilt parameter and score in Task 2 becomes stronger, whereas if participants focus shorter on the well spacing, there is a weaker relationship between the dwell time on the tilt parameter and the score in Task 2.

\section{Discussion}


\subsection{Task-solving performance of experts and novices} 
In this work, we compared the performance of experts and novices when they solve problems using Quantum Composer. The participants were recruited from students who completed at least one lecture of quantum mechanics and from PhD's and postdocs who are specialized in quantum technologies. With this selection, we expected to be able to compare the performance of experts and novices. Indeed, we were able to observe significant differences in the pretest score and assigned the participants accordingly to the groups of experts and novices; the threshold was at a pretest score of 0.5. Similar to the pretest, the experts reached a higher number of correct statements during the reasoning tasks and the lower number of incorrect statements, leading to a higher total score. The experts also reached a higher score in both Tasks 1 and 2 separately. To understand how participants apply the content knowledge during Quantum Composer tasks, i.e., how do they relate the content knowledge to the information in the external representations in Quantum Composer, we evaluated the number of relevant concepts the participants used during the reasoning. Despite the clear difference in the performance between experts and novices, we did not find a direct linear relation between the pretest and the total score. Instead, we found a significant mediation of the number of concepts between the pretest and the total score. This means that only those participants who were able to relate the content knowledge to the information depicted in the representations and formulate it were better able to solve problems using Quantum Composer. As mentioned above, we use the number of concepts mentioned during the reasoning tasks as a measure for the metarepresentational competence of participants. One possible interpretation of this finding is that additional to the content knowledge in quantum mechanics, the participants also need metarepresentational competence to solve problems with Quantum Composer successfully. This suggests the importance to explicitly make students relate various representations  in the Quantum Composer to prior knowledge, e.g. using think-aloud techniques, in order to become true experts. These results are in line with the suggestions of Wawro et al. who suggested to provide scaffolding support for students when solving problems or learning quantum mechanics \cite{wawro2020students}. In our study, Quantum Composer is an environment in which students need to integrate information from MERs to solve problems. In this sense, it is not unique but comparable to several other simulation environments, and we also believe that our findings are not restricted to problems in quantum mechanics that are related to Quantum Composer but also apply to those multimedia learning environments in which learners need to extract and integrate information from MERs to solve problems successfully. 

\subsection{Gaze behavior of experts and novices}
\subsubsection{Visual attention on relevant and irrelevant areas}
We found a clear difference in the gaze data of experts and novices. In both Tasks 1 and 2, it was necessary to extract information from the graph that displays the energy states, the wave function, and the double-well potential, all while modifying the well spacing and the potential height. It was not necessary to vary the tilt parameter of the potential or pay attention to the functional form of potential. In both Quantum Composer tasks, we found that experts focus relatively longer on the well spacing and the potential height parameter and clearly less on the tilt parameter. This difference in the relative dwell time is also noticeable in the gaze patterns of the best and the lowest performers: Experts clearly focus on the relevant areas and pay very little attention to irrelevant elements. It is also noticeable that the experts even have a lower number of fixations on the graph, which indicates that they are more efficient in extracting the information. All of these observations confirm the information reduction hypothesis, which states that experts are more efficient in extracting information, and focus longer on relevant and shorter on irrelevant areas \cite{gegenfurtner2011expertise}. 

\subsubsection{Visual attention on the graph}
Apart from that, it is important to notice that for both groups, experts and novices, the graph receives by far the most visual attention among all relevant and irrelevant areas; nearly 80 \% of the entire visual attention on the Quantum Composer interface is on the graph during the solution of each of the two tasks. This finding implies that both groups noticed the importance of the graph in solving the problem. At this point we cannot say whether the participants also paid a lot of attention to the graph because it is a very familiar representation for them. It is also important to realize that the information that had to be extracted from the graph, namely the numerical value of the energy splitting between the ground state and the first excited state, was redundant to the information of the energy difference between these two states displayed at the top right of Quantum Composer interface. However, among the 36 participants, no one used this number to solve Task 1 or Task 2 even though it might be easier to extract because it is only a single number in comparison to the two visual energy levels in the graph that need to be compared for the solution. 
\subsubsection{Relation between visual attention and performance}
Considering all these findings in Task 2, we found that there is a linear relationship between the score in Task 2 and the total dwell time on the graph as well as the total dwell time on the tilt parameter, whereas the latter quantity is moderated by the total dwell time on the well spacing. This finding implies that it is not only important to identify relevant information but also to identify and disregard irrelevant information to achieve a high score during problem-solving with Quantum Composer. 
From a perspective of representational competence, these findings indicate that there is a difference in visual fluency between experts and novices \cite{rau2017make}. Apart from visual fluency, there is also a difference in the fluency in using Quantum Composer, which is visible in the number of modifications of the parameters. For instance, in both Tasks 1 and 2, novices modify the scaling of the graph more often than experts, and they have a higher total number of modifications of all parameters. In both tasks, it was not required to change the scaling of the graph because the relationship between the dependent and independent variables are also visible without any changes to the scaling. Thus, re-scaling the graph is unnecessary to solve the problem.

Overall, when designing a multimedia environment with MERs that contain content-relevant information or when integrating such environments in the class, we suggest that students receive an additional instruction for the MERs implemented in the learning environment or scaffolding support to ensure optimal learning outcomes. Furthermore, we recommend to discuss relevant and irrelevant information in exemplary problems. Such a discussion may help enhancing students visual fluency \cite{rau2017make}, so that they require less effort during the solution and solve problems and learn more effectively. This process to enhance students' visual fluency may also be supported by eye-movement modeling examples of experts \cite{gegenfurtner2017effects}.

\section{Conclusion}
In this study, we investigated how experts and novices solve tasks in quantum mechanics using Quantum Composer, a software that allows instructors to set up custom simulation environments in which users to visually simulate quantum states, in which participants have to extract and integrate information from various types of MERs. After a pretest, the participants were divided into groups of experts and novices in quantum mechanics. Then, they solved reasoning tasks using Quantum Composer and nearly all students completed at least the first two tasks. In both tasks, the experts performed significantly better than novices. We found that there was no direct effect between the pretest score and the total score. Instead, we found a significant indirect effect between the two variables with the number of relevant concepts mentioned during the reasoning tasks as the mediator. This finding points towards the importance of meta-representational competence additionally to content knowledge when learning or problem-solving with interactive visualization apps. Furthermore, we found that there is a clear difference in the gaze behavior between experts and novices. Experts focus longer on relevant and shorter on irrelevant areas of Quantum Composer, which confirms the information-reduction hypothesis. Overall, this study contributes to the understanding of the differences between experts and novices problem-solving with interactive visualization tools, and identifies the requirements for successful learning with such tools.

For the development of interactive visualization tools, we recommend to (\textit{a}) be aware of the representational affordances of each representation added to the interface as students may be unfamiliar to them, (\textit{b}) potentially be flexible to provide additional scaffolding support for students to extract and integrate information (such as visual highlights of relevant elements), and (\textit{c}) to code the same information in different types of representation, such as graphs, text, equation and schematic visualizations. In this way, the interactive visualization tool would be able to compensate shortcomings in students' representational competences. 

For a usage of Quantum Composer in class, we advise teachers  (\textit{i}) to introduce each representation in the specific context before using the tool, (\textit{ii}), potentially provide visual aids to support information identification, extraction and integration from representations, (\textit{iii}) demonstrate beforehand how to reason with the information depicted in representations, and (\textit{iv}) use simple training tasks to get familiar with the features and the interface of Quantum Composer. 

Apart from that, it is important to be aware of the limitation of this work that we only studied the task-solving process of experts and novices, consisting of identification, extraction and evaluation of information as well as reasoning based on evidence, with multiple representations embedded in a static interface of Quantum Composer. However, tools such as Quantum Composer are much richer in their features and fields of application in class. To obtain a more complete overview of the applications of interactive visualization tools in class, future research should address, for instance, the interactive design process of visualization tools such as Quantum Composer. The elements in the static version of Quantum Composer used here may not have been intuitive and most effective for some students. On one hand, a self-designed interface may therefore be more beneficial for learning for some students, because students may arrange elements, delete irrelevant or unhelpful elements and include other helpful information, and optimize their learning experience and performance in this way. On the other hand, the interface design process also requires more cognitive resources and awareness of the task demands as well as how to optimize the own learning process and task performance. 
Another limitation of this work is the assignment of expertise levels based on previous knowledge. Previously, several parameters have been used to discriminate between experts and novices. For instance, Dogusoy-Taylan and Cagiltay identified expertise levels based on their profession \cite{dogusoy2014cognitive},  graph literacy when solving graph problems \cite{okan2016people}, or spatial working memory capacity \cite{huang2016gender}.  Here, we are in line with previous works by Ho et al. (2014) and Richter et al. (2021) who also determine expertise levels based on prior domain knowledge. \cite{ho2014prior,richter2021poor}. However, the results showed that there is no direct effect between expertise levels (measured in terms of pretest score) and Quantum Composer task performance which might be indication that there are other measures than prior knowledge that may explain expertise levels in Quantum Composer tasks. 

In conclusion, this work provides essential information of the task-solving process with an interactive visualization tool, but relevant questions for a broader range of applications of interaction visualization tools in class need to be addressed in the future.



\bibliography{bibliography.bib}

\begin{thebibliography}{67}%
\makeatletter
\providecommand \@ifxundefined [1]{%
 \@ifx{#1\undefined}
}%
\providecommand \@ifnum [1]{%
 \ifnum #1\expandafter \@firstoftwo
 \else \expandafter \@secondoftwo
 \fi
}%
\providecommand \@ifx [1]{%
 \ifx #1\expandafter \@firstoftwo
 \else \expandafter \@secondoftwo
 \fi
}%
\providecommand \natexlab [1]{#1}%
\providecommand \enquote  [1]{``#1''}%
\providecommand \bibnamefont  [1]{#1}%
\providecommand \bibfnamefont [1]{#1}%
\providecommand \citenamefont [1]{#1}%
\providecommand \href@noop [0]{\@secondoftwo}%
\providecommand \href [0]{\begingroup \@sanitize@url \@href}%
\providecommand \@href[1]{\@@startlink{#1}\@@href}%
\providecommand \@@href[1]{\endgroup#1\@@endlink}%
\providecommand \@sanitize@url [0]{\catcode `\\12\catcode `\$12\catcode
  `\&12\catcode `\#12\catcode `\^12\catcode `\_12\catcode `\%12\relax}%
\providecommand \@@startlink[1]{}%
\providecommand \@@endlink[0]{}%
\providecommand \url  [0]{\begingroup\@sanitize@url \@url }%
\providecommand \@url [1]{\endgroup\@href {#1}{\urlprefix }}%
\providecommand \urlprefix  [0]{URL }%
\providecommand \Eprint [0]{\href }%
\providecommand \doibase [0]{https://doi.org/}%
\providecommand \selectlanguage [0]{\@gobble}%
\providecommand \bibinfo  [0]{\@secondoftwo}%
\providecommand \bibfield  [0]{\@secondoftwo}%
\providecommand \translation [1]{[#1]}%
\providecommand \BibitemOpen [0]{}%
\providecommand \bibitemStop [0]{}%
\providecommand \bibitemNoStop [0]{.\EOS\space}%
\providecommand \EOS [0]{\spacefactor3000\relax}%
\providecommand \BibitemShut  [1]{\csname bibitem#1\endcsname}%
\let\auto@bib@innerbib\@empty
\bibitem [{\citenamefont {Fox}\ \emph {et~al.}(2020)\citenamefont {Fox},
  \citenamefont {Zwickl},\ and\ \citenamefont
  {Lewandowski}}]{fox2020preparing}%
  \BibitemOpen
  \bibfield  {author} {\bibinfo {author} {\bibfnamefont {M.~F.}\ \bibnamefont
  {Fox}}, \bibinfo {author} {\bibfnamefont {B.~M.}\ \bibnamefont {Zwickl}},\
  and\ \bibinfo {author} {\bibfnamefont {H.}~\bibnamefont {Lewandowski}},\
  }\href@noop {} {\bibfield  {journal} {\bibinfo  {journal} {Physical Review
  Physics Education Research}\ }\textbf {\bibinfo {volume} {16}},\ \bibinfo
  {pages} {020131} (\bibinfo {year} {2020})}\BibitemShut {NoStop}%
\bibitem [{\citenamefont {Larkin}\ \emph {et~al.}(1980)\citenamefont {Larkin},
  \citenamefont {McDermott}, \citenamefont {Simon},\ and\ \citenamefont
  {Simon}}]{larkin1980expert}%
  \BibitemOpen
  \bibfield  {author} {\bibinfo {author} {\bibfnamefont {J.}~\bibnamefont
  {Larkin}}, \bibinfo {author} {\bibfnamefont {J.}~\bibnamefont {McDermott}},
  \bibinfo {author} {\bibfnamefont {D.~P.}\ \bibnamefont {Simon}},\ and\
  \bibinfo {author} {\bibfnamefont {H.~A.}\ \bibnamefont {Simon}},\ }\href@noop
  {} {\bibfield  {journal} {\bibinfo  {journal} {Science}\ }\textbf {\bibinfo
  {volume} {208}},\ \bibinfo {pages} {1335} (\bibinfo {year}
  {1980})}\BibitemShut {NoStop}%
\bibitem [{\citenamefont {Scherer}\ \emph {et~al.}(2019)\citenamefont
  {Scherer}, \citenamefont {Siddiq},\ and\ \citenamefont
  {Tondeur}}]{scherer2019technology}%
  \BibitemOpen
  \bibfield  {author} {\bibinfo {author} {\bibfnamefont {R.}~\bibnamefont
  {Scherer}}, \bibinfo {author} {\bibfnamefont {F.}~\bibnamefont {Siddiq}},\
  and\ \bibinfo {author} {\bibfnamefont {J.}~\bibnamefont {Tondeur}},\
  }\href@noop {} {\bibfield  {journal} {\bibinfo  {journal} {Computers \&
  Education}\ }\textbf {\bibinfo {volume} {128}},\ \bibinfo {pages} {13}
  (\bibinfo {year} {2019})}\BibitemShut {NoStop}%
\bibitem [{\citenamefont {Fredlund}\ \emph {et~al.}(2014)\citenamefont
  {Fredlund}, \citenamefont {Linder}, \citenamefont {Airey},\ and\
  \citenamefont {Linder}}]{fredlund2014unpacking}%
  \BibitemOpen
  \bibfield  {author} {\bibinfo {author} {\bibfnamefont {T.}~\bibnamefont
  {Fredlund}}, \bibinfo {author} {\bibfnamefont {C.}~\bibnamefont {Linder}},
  \bibinfo {author} {\bibfnamefont {J.}~\bibnamefont {Airey}},\ and\ \bibinfo
  {author} {\bibfnamefont {A.}~\bibnamefont {Linder}},\ }\href@noop {}
  {\bibfield  {journal} {\bibinfo  {journal} {Physical Review Special
  Topics-Physics Education Research}\ }\textbf {\bibinfo {volume} {10}},\
  \bibinfo {pages} {020129} (\bibinfo {year} {2014})}\BibitemShut {NoStop}%
\bibitem [{\citenamefont {K{\"u}chemann}\ \emph
  {et~al.}(2020{\natexlab{a}})\citenamefont {K{\"u}chemann}, \citenamefont
  {Klein}, \citenamefont {Fouckhardt}, \citenamefont {Gr{\"o}ber},\ and\
  \citenamefont {Kuhn}}]{kuchemann2020students}%
  \BibitemOpen
  \bibfield  {author} {\bibinfo {author} {\bibfnamefont {S.}~\bibnamefont
  {K{\"u}chemann}}, \bibinfo {author} {\bibfnamefont {P.}~\bibnamefont
  {Klein}}, \bibinfo {author} {\bibfnamefont {H.}~\bibnamefont {Fouckhardt}},
  \bibinfo {author} {\bibfnamefont {S.}~\bibnamefont {Gr{\"o}ber}},\ and\
  \bibinfo {author} {\bibfnamefont {J.}~\bibnamefont {Kuhn}},\ }\href@noop {}
  {\bibfield  {journal} {\bibinfo  {journal} {Physical Review Physics Education
  Research}\ }\textbf {\bibinfo {volume} {16}},\ \bibinfo {pages} {010112}
  (\bibinfo {year} {2020}{\natexlab{a}})}\BibitemShut {NoStop}%
\bibitem [{\citenamefont {Klein}\ \emph {et~al.}(2019)\citenamefont {Klein},
  \citenamefont {K{\"u}chemann}, \citenamefont {Br{\"u}ckner}, \citenamefont
  {Zlatkin-Troitschanskaia},\ and\ \citenamefont {Kuhn}}]{klein2019student}%
  \BibitemOpen
  \bibfield  {author} {\bibinfo {author} {\bibfnamefont {P.}~\bibnamefont
  {Klein}}, \bibinfo {author} {\bibfnamefont {S.}~\bibnamefont
  {K{\"u}chemann}}, \bibinfo {author} {\bibfnamefont {S.}~\bibnamefont
  {Br{\"u}ckner}}, \bibinfo {author} {\bibfnamefont {O.}~\bibnamefont
  {Zlatkin-Troitschanskaia}},\ and\ \bibinfo {author} {\bibfnamefont
  {J.}~\bibnamefont {Kuhn}},\ }\href@noop {} {\bibfield  {journal} {\bibinfo
  {journal} {Physical Review Physics Education Research}\ }\textbf {\bibinfo
  {volume} {15}},\ \bibinfo {pages} {020116} (\bibinfo {year}
  {2019})}\BibitemShut {NoStop}%
\bibitem [{\citenamefont {Zaman~Ahmed}\ \emph {et~al.}(2021)\citenamefont
  {Zaman~Ahmed}, \citenamefont {Jensen}, \citenamefont {Weidner}, \citenamefont
  {S{\o}rensen}, \citenamefont {Mudrich},\ and\ \citenamefont
  {Sherson}}]{zaman2021quantum}%
  \BibitemOpen
  \bibfield  {author} {\bibinfo {author} {\bibfnamefont {S.}~\bibnamefont
  {Zaman~Ahmed}}, \bibinfo {author} {\bibfnamefont {J.~H.~M.}\ \bibnamefont
  {Jensen}}, \bibinfo {author} {\bibfnamefont {C.~A.}\ \bibnamefont {Weidner}},
  \bibinfo {author} {\bibfnamefont {J.~J.}\ \bibnamefont {S{\o}rensen}},
  \bibinfo {author} {\bibfnamefont {M.}~\bibnamefont {Mudrich}},\ and\ \bibinfo
  {author} {\bibfnamefont {J.~F.}\ \bibnamefont {Sherson}},\ }\href@noop {}
  {\bibfield  {journal} {\bibinfo  {journal} {American Journal of Physics}\
  }\textbf {\bibinfo {volume} {89}},\ \bibinfo {pages} {307} (\bibinfo {year}
  {2021})}\BibitemShut {NoStop}%
\bibitem [{\citenamefont {Weidner}\ \emph {et~al.}(2020)\citenamefont
  {Weidner}, \citenamefont {Ahmed}, \citenamefont {Jensen}, \citenamefont
  {Sherson},\ and\ \citenamefont {Lewandowski}}]{Carrie_2020}%
  \BibitemOpen
  \bibfield  {author} {\bibinfo {author} {\bibfnamefont {C.~A.}\ \bibnamefont
  {Weidner}}, \bibinfo {author} {\bibfnamefont {S.~Z.}\ \bibnamefont {Ahmed}},
  \bibinfo {author} {\bibfnamefont {J.~H.~M.}\ \bibnamefont {Jensen}}, \bibinfo
  {author} {\bibfnamefont {J.~F.}\ \bibnamefont {Sherson}},\ and\ \bibinfo
  {author} {\bibfnamefont {H.~J.}\ \bibnamefont {Lewandowski}},\ }in\
  \href@noop {} {\emph {\bibinfo {booktitle} {Physics Education Research
  Conference 2020}}},\ \bibinfo {series and number} {PER Conference}\ (\bibinfo
  {address} {Virtual Conference},\ \bibinfo {year} {2020})\ pp.\ \bibinfo
  {pages} {563--568}\BibitemShut {NoStop}%
\bibitem [{QTE()}]{QTEdu}%
  \BibitemOpen
  \href@noop {} {\bibinfo {title} {{QTEdu, Coordination and Support Action for
  Quantum Technology Education. Quantum Flagship.}}},\ \bibinfo {howpublished}
  {\url{https://qtedu.eu/}},\ \bibinfo {note} {accessed:
  2022-03-15}\BibitemShut {NoStop}%
\bibitem [{Dig()}]{DigCompEdu}%
  \BibitemOpen
  \href@noop {} {\bibinfo {title} {Digital competence framework for educators
  (digcompedu). eu science hub - european commission.}}\BibitemShut {Stop}%
\bibitem [{\citenamefont {Gehrke}\ \emph {et~al.}(2020)\citenamefont {Gehrke},
  \citenamefont {Schauss}, \citenamefont {K{\"u}sters},\ and\ \citenamefont
  {Gries}}]{gehrke2020experiencing}%
  \BibitemOpen
  \bibfield  {author} {\bibinfo {author} {\bibfnamefont {I.}~\bibnamefont
  {Gehrke}}, \bibinfo {author} {\bibfnamefont {M.}~\bibnamefont {Schauss}},
  \bibinfo {author} {\bibfnamefont {D.}~\bibnamefont {K{\"u}sters}},\ and\
  \bibinfo {author} {\bibfnamefont {T.}~\bibnamefont {Gries}},\ }\href@noop {}
  {\bibfield  {journal} {\bibinfo  {journal} {Procedia Manufacturing}\ }\textbf
  {\bibinfo {volume} {45}},\ \bibinfo {pages} {177} (\bibinfo {year}
  {2020})}\BibitemShut {NoStop}%
\bibitem [{\citenamefont {Passante}\ and\ \citenamefont
  {Kohnle}(2019)}]{passante2019enhancing}%
  \BibitemOpen
  \bibfield  {author} {\bibinfo {author} {\bibfnamefont {G.}~\bibnamefont
  {Passante}}\ and\ \bibinfo {author} {\bibfnamefont {A.}~\bibnamefont
  {Kohnle}},\ }\href@noop {} {\bibfield  {journal} {\bibinfo  {journal}
  {Physical Review Physics Education Research}\ }\textbf {\bibinfo {volume}
  {15}},\ \bibinfo {pages} {010110} (\bibinfo {year} {2019})}\BibitemShut
  {NoStop}%
\bibitem [{\citenamefont {Emigh}\ \emph {et~al.}(2015)\citenamefont {Emigh},
  \citenamefont {Passante},\ and\ \citenamefont {Shaffer}}]{emigh2015student}%
  \BibitemOpen
  \bibfield  {author} {\bibinfo {author} {\bibfnamefont {P.~J.}\ \bibnamefont
  {Emigh}}, \bibinfo {author} {\bibfnamefont {G.}~\bibnamefont {Passante}},\
  and\ \bibinfo {author} {\bibfnamefont {P.~S.}\ \bibnamefont {Shaffer}},\
  }\href@noop {} {\bibfield  {journal} {\bibinfo  {journal} {Physical Review
  Special Topics-Physics Education Research}\ }\textbf {\bibinfo {volume}
  {11}},\ \bibinfo {pages} {020112} (\bibinfo {year} {2015})}\BibitemShut
  {NoStop}%
\bibitem [{\citenamefont {McKagan}\ \emph
  {et~al.}(2008{\natexlab{a}})\citenamefont {McKagan}, \citenamefont
  {Perkins},\ and\ \citenamefont {Wieman}}]{mckagan2008deeper}%
  \BibitemOpen
  \bibfield  {author} {\bibinfo {author} {\bibfnamefont {S.}~\bibnamefont
  {McKagan}}, \bibinfo {author} {\bibfnamefont {K.}~\bibnamefont {Perkins}},\
  and\ \bibinfo {author} {\bibfnamefont {C.}~\bibnamefont {Wieman}},\
  }\href@noop {} {\bibfield  {journal} {\bibinfo  {journal} {Physical Review
  Special Topics-Physics Education Research}\ }\textbf {\bibinfo {volume}
  {4}},\ \bibinfo {pages} {020103} (\bibinfo {year}
  {2008}{\natexlab{a}})}\BibitemShut {NoStop}%
\bibitem [{\citenamefont {Wittmann}\ \emph {et~al.}(2005)\citenamefont
  {Wittmann}, \citenamefont {Morgan},\ and\ \citenamefont
  {Bao}}]{wittmann2005addressing}%
  \BibitemOpen
  \bibfield  {author} {\bibinfo {author} {\bibfnamefont {M.~C.}\ \bibnamefont
  {Wittmann}}, \bibinfo {author} {\bibfnamefont {J.~T.}\ \bibnamefont
  {Morgan}},\ and\ \bibinfo {author} {\bibfnamefont {L.}~\bibnamefont {Bao}},\
  }\href@noop {} {\bibfield  {journal} {\bibinfo  {journal} {European Journal
  of Physics}\ }\textbf {\bibinfo {volume} {26}},\ \bibinfo {pages} {939}
  (\bibinfo {year} {2005})}\BibitemShut {NoStop}%
\bibitem [{\citenamefont {Morgan}\ and\ \citenamefont
  {Wittmann}(2006)}]{morgan2006examining}%
  \BibitemOpen
  \bibfield  {author} {\bibinfo {author} {\bibfnamefont {J.~T.}\ \bibnamefont
  {Morgan}}\ and\ \bibinfo {author} {\bibfnamefont {M.~C.}\ \bibnamefont
  {Wittmann}},\ }in\ \href@noop {} {\emph {\bibinfo {booktitle} {AIP Conference
  Proceedings}}},\ Vol.\ \bibinfo {volume} {818}\ (\bibinfo {organization}
  {American Institute of Physics},\ \bibinfo {year} {2006})\ pp.\ \bibinfo
  {pages} {73--76}\BibitemShut {NoStop}%
\bibitem [{\citenamefont {Griffith}(1995)}]{Griffith_1995}%
  \BibitemOpen
  \bibfield  {author} {\bibinfo {author} {\bibfnamefont {D.~J.}\ \bibnamefont
  {Griffith}},\ }\href@noop {} {\emph {\bibinfo {title} {Introduction to
  Quantum Mechanics}}}\ (\bibinfo  {publisher} {Pearson Education},\ \bibinfo
  {year} {1995})\BibitemShut {NoStop}%
\bibitem [{\citenamefont {Griffith}\ \emph {et~al.}(1977)\citenamefont
  {Griffith}, \citenamefont {Diu},\ and\ \citenamefont
  {Lalo\:{e}}}]{Cohen-Tannoudji_1977}%
  \BibitemOpen
  \bibfield  {author} {\bibinfo {author} {\bibfnamefont {D.~J.}\ \bibnamefont
  {Griffith}}, \bibinfo {author} {\bibfnamefont {B.}~\bibnamefont {Diu}},\ and\
  \bibinfo {author} {\bibfnamefont {F.}~\bibnamefont {Lalo\:{e}}},\ }\href@noop
  {} {\emph {\bibinfo {title} {Quantum Mechanics}}}\ (\bibinfo  {publisher}
  {Wiley},\ \bibinfo {year} {1977})\BibitemShut {NoStop}%
\bibitem [{\citenamefont {Shankar}(2011)}]{Shankar_2011}%
  \BibitemOpen
  \bibfield  {author} {\bibinfo {author} {\bibfnamefont {R.}~\bibnamefont
  {Shankar}},\ }\href@noop {} {\emph {\bibinfo {title} {Principles of Quantum
  Mechanics (2nd ed.)}}}\ (\bibinfo  {publisher} {Plenum Press},\ \bibinfo
  {year} {2011})\BibitemShut {NoStop}%
\bibitem [{\citenamefont {Stadermann}\ \emph {et~al.}(2019)\citenamefont
  {Stadermann}, \citenamefont {van~den Berg},\ and\ \citenamefont
  {Goedhart}}]{stadermann2019analysis}%
  \BibitemOpen
  \bibfield  {author} {\bibinfo {author} {\bibfnamefont {H.}~\bibnamefont
  {Stadermann}}, \bibinfo {author} {\bibfnamefont {E.}~\bibnamefont {van~den
  Berg}},\ and\ \bibinfo {author} {\bibfnamefont {M.}~\bibnamefont
  {Goedhart}},\ }\href@noop {} {\bibfield  {journal} {\bibinfo  {journal}
  {Physical Review Physics Education Research}\ }\textbf {\bibinfo {volume}
  {15}},\ \bibinfo {pages} {010130} (\bibinfo {year} {2019})}\BibitemShut
  {NoStop}%
\bibitem [{\citenamefont {Aiello}\ \emph {et~al.}(2021)\citenamefont {Aiello},
  \citenamefont {Awschalom}, \citenamefont {Bernien}, \citenamefont {Brower},
  \citenamefont {Brown}, \citenamefont {Brun}, \citenamefont {Caram},
  \citenamefont {Chitambar}, \citenamefont {Di~Felice}, \citenamefont {Edmonds}
  \emph {et~al.}}]{aiello2021achieving}%
  \BibitemOpen
  \bibfield  {author} {\bibinfo {author} {\bibfnamefont {C.~D.}\ \bibnamefont
  {Aiello}}, \bibinfo {author} {\bibfnamefont {D.}~\bibnamefont {Awschalom}},
  \bibinfo {author} {\bibfnamefont {H.}~\bibnamefont {Bernien}}, \bibinfo
  {author} {\bibfnamefont {T.}~\bibnamefont {Brower}}, \bibinfo {author}
  {\bibfnamefont {K.~R.}\ \bibnamefont {Brown}}, \bibinfo {author}
  {\bibfnamefont {T.~A.}\ \bibnamefont {Brun}}, \bibinfo {author}
  {\bibfnamefont {J.~R.}\ \bibnamefont {Caram}}, \bibinfo {author}
  {\bibfnamefont {E.}~\bibnamefont {Chitambar}}, \bibinfo {author}
  {\bibfnamefont {R.}~\bibnamefont {Di~Felice}}, \bibinfo {author}
  {\bibfnamefont {K.~M.}\ \bibnamefont {Edmonds}}, \emph {et~al.},\ }\href@noop
  {} {\bibfield  {journal} {\bibinfo  {journal} {Quantum Science and
  Technology}\ }\textbf {\bibinfo {volume} {6}},\ \bibinfo {pages} {030501}
  (\bibinfo {year} {2021})}\BibitemShut {NoStop}%
\bibitem [{\citenamefont {Singh}\ and\ \citenamefont
  {Marshman}(2015)}]{singh2015review}%
  \BibitemOpen
  \bibfield  {author} {\bibinfo {author} {\bibfnamefont {C.}~\bibnamefont
  {Singh}}\ and\ \bibinfo {author} {\bibfnamefont {E.}~\bibnamefont
  {Marshman}},\ }\href@noop {} {\bibfield  {journal} {\bibinfo  {journal}
  {Physical Review Special Topics-Physics Education Research}\ }\textbf
  {\bibinfo {volume} {11}},\ \bibinfo {pages} {020117} (\bibinfo {year}
  {2015})}\BibitemShut {NoStop}%
\bibitem [{\citenamefont {Krijtenburg-Lewerissa}\ \emph
  {et~al.}(2017)\citenamefont {Krijtenburg-Lewerissa}, \citenamefont {Pol},
  \citenamefont {Brinkman},\ and\ \citenamefont
  {Van~Joolingen}}]{krijtenburg2017insights}%
  \BibitemOpen
  \bibfield  {author} {\bibinfo {author} {\bibfnamefont {K.}~\bibnamefont
  {Krijtenburg-Lewerissa}}, \bibinfo {author} {\bibfnamefont {H.~J.}\
  \bibnamefont {Pol}}, \bibinfo {author} {\bibfnamefont {A.}~\bibnamefont
  {Brinkman}},\ and\ \bibinfo {author} {\bibfnamefont {W.}~\bibnamefont
  {Van~Joolingen}},\ }\href@noop {} {\bibfield  {journal} {\bibinfo  {journal}
  {Physical review physics education research}\ }\textbf {\bibinfo {volume}
  {13}},\ \bibinfo {pages} {010109} (\bibinfo {year} {2017})}\BibitemShut
  {NoStop}%
\bibitem [{\citenamefont {Cataloglu}\ and\ \citenamefont
  {Robinett}(2002)}]{cataloglu2002testing}%
  \BibitemOpen
  \bibfield  {author} {\bibinfo {author} {\bibfnamefont {E.}~\bibnamefont
  {Cataloglu}}\ and\ \bibinfo {author} {\bibfnamefont {R.~W.}\ \bibnamefont
  {Robinett}},\ }\href@noop {} {\bibfield  {journal} {\bibinfo  {journal}
  {American Journal of Physics}\ }\textbf {\bibinfo {volume} {70}},\ \bibinfo
  {pages} {238} (\bibinfo {year} {2002})}\BibitemShut {NoStop}%
\bibitem [{\citenamefont {Kohnle}\ \emph {et~al.}(2018)\citenamefont {Kohnle},
  \citenamefont {Benfield}, \citenamefont {Cassettari}, \citenamefont
  {Edwards}, \citenamefont {Fomins}, \citenamefont {Gillies}, \citenamefont
  {H{\"a}hner}, \citenamefont {Hooley}, \citenamefont {Korolkova},
  \citenamefont {Llama} \emph {et~al.}}]{kohnle2018quvis}%
  \BibitemOpen
  \bibfield  {author} {\bibinfo {author} {\bibfnamefont {A.}~\bibnamefont
  {Kohnle}}, \bibinfo {author} {\bibfnamefont {C.}~\bibnamefont {Benfield}},
  \bibinfo {author} {\bibfnamefont {D.}~\bibnamefont {Cassettari}}, \bibinfo
  {author} {\bibfnamefont {T.}~\bibnamefont {Edwards}}, \bibinfo {author}
  {\bibfnamefont {A.}~\bibnamefont {Fomins}}, \bibinfo {author} {\bibfnamefont
  {A.}~\bibnamefont {Gillies}}, \bibinfo {author} {\bibfnamefont
  {G.}~\bibnamefont {H{\"a}hner}}, \bibinfo {author} {\bibfnamefont
  {C.}~\bibnamefont {Hooley}}, \bibinfo {author} {\bibfnamefont
  {N.}~\bibnamefont {Korolkova}}, \bibinfo {author} {\bibfnamefont
  {J.}~\bibnamefont {Llama}}, \emph {et~al.},\ }\href@noop {} {\bibinfo {title}
  {Quvis: The quantum mechanics visualization project}} (\bibinfo {year}
  {2018})\BibitemShut {NoStop}%
\bibitem [{\citenamefont {Kohnle}\ \emph {et~al.}(2010)\citenamefont {Kohnle},
  \citenamefont {Douglass}, \citenamefont {Edwards}, \citenamefont {Gillies},
  \citenamefont {Hooley},\ and\ \citenamefont
  {Sinclair}}]{kohnle2010developing}%
  \BibitemOpen
  \bibfield  {author} {\bibinfo {author} {\bibfnamefont {A.}~\bibnamefont
  {Kohnle}}, \bibinfo {author} {\bibfnamefont {M.}~\bibnamefont {Douglass}},
  \bibinfo {author} {\bibfnamefont {T.~J.}\ \bibnamefont {Edwards}}, \bibinfo
  {author} {\bibfnamefont {A.~D.}\ \bibnamefont {Gillies}}, \bibinfo {author}
  {\bibfnamefont {C.~A.}\ \bibnamefont {Hooley}},\ and\ \bibinfo {author}
  {\bibfnamefont {B.~D.}\ \bibnamefont {Sinclair}},\ }\href@noop {} {\bibfield
  {journal} {\bibinfo  {journal} {European Journal of Physics}\ }\textbf
  {\bibinfo {volume} {31}},\ \bibinfo {pages} {1441} (\bibinfo {year}
  {2010})}\BibitemShut {NoStop}%
\bibitem [{\citenamefont {S{\o}rensen}\ \emph {et~al.}(2019)\citenamefont
  {S{\o}rensen}, \citenamefont {Jensen}, \citenamefont {Heinzel},\ and\
  \citenamefont {Sherson}}]{sorensen2019qengine}%
  \BibitemOpen
  \bibfield  {author} {\bibinfo {author} {\bibfnamefont {J.~J.}\ \bibnamefont
  {S{\o}rensen}}, \bibinfo {author} {\bibfnamefont {J.}~\bibnamefont {Jensen}},
  \bibinfo {author} {\bibfnamefont {T.}~\bibnamefont {Heinzel}},\ and\ \bibinfo
  {author} {\bibfnamefont {J.~F.}\ \bibnamefont {Sherson}},\ }\href@noop {}
  {\bibfield  {journal} {\bibinfo  {journal} {Computer Physics Communications}\
  }\textbf {\bibinfo {volume} {243}},\ \bibinfo {pages} {135} (\bibinfo {year}
  {2019})}\BibitemShut {NoStop}%
\bibitem [{\citenamefont {Weidner}\ \emph {et~al.}()\citenamefont {Weidner},
  \citenamefont {Ahmed}, \citenamefont {Jensen},\ and\ \citenamefont
  {Sherson}}]{quatomic}%
  \BibitemOpen
  \bibfield  {author} {\bibinfo {author} {\bibfnamefont {C.~A.}\ \bibnamefont
  {Weidner}}, \bibinfo {author} {\bibfnamefont {S.~Z.}\ \bibnamefont {Ahmed}},
  \bibinfo {author} {\bibfnamefont {J.~H.~M.}\ \bibnamefont {Jensen}},\ and\
  \bibinfo {author} {\bibfnamefont {J.~F.}\ \bibnamefont {Sherson}},\
  }\href@noop {} {\bibinfo {title} {Publications using quatomic software}},\
  \bibinfo {howpublished} {\url{https://www.quatomic.com/composer/}},\ \bibinfo
  {note} {retrieved 3/25/2021}\BibitemShut {NoStop}%
\bibitem [{\citenamefont {Johansson}\ \emph {et~al.}(2012)\citenamefont
  {Johansson}, \citenamefont {Nation},\ and\ \citenamefont
  {Nori}}]{johansson2012qutip}%
  \BibitemOpen
  \bibfield  {author} {\bibinfo {author} {\bibfnamefont {J.~R.}\ \bibnamefont
  {Johansson}}, \bibinfo {author} {\bibfnamefont {P.~D.}\ \bibnamefont
  {Nation}},\ and\ \bibinfo {author} {\bibfnamefont {F.}~\bibnamefont {Nori}},\
  }\href@noop {} {\bibfield  {journal} {\bibinfo  {journal} {Computer Physics
  Communications}\ }\textbf {\bibinfo {volume} {183}},\ \bibinfo {pages} {1760}
  (\bibinfo {year} {2012})}\BibitemShut {NoStop}%
\bibitem [{\citenamefont {Babayev}\ \emph {et~al.}(2021)\citenamefont
  {Babayev}, \citenamefont {Andersson}, \citenamefont {Dalmau}, \citenamefont
  {Im},\ and\ \citenamefont {Johansson}}]{babayev2021computational}%
  \BibitemOpen
  \bibfield  {author} {\bibinfo {author} {\bibfnamefont {R.}~\bibnamefont
  {Babayev}}, \bibinfo {author} {\bibfnamefont {A.}~\bibnamefont {Andersson}},
  \bibinfo {author} {\bibfnamefont {A.~S.}\ \bibnamefont {Dalmau}}, \bibinfo
  {author} {\bibfnamefont {H.~G.}\ \bibnamefont {Im}},\ and\ \bibinfo {author}
  {\bibfnamefont {B.}~\bibnamefont {Johansson}},\ }\href@noop {} {\bibfield
  {journal} {\bibinfo  {journal} {International Journal of Hydrogen Energy}\
  }\textbf {\bibinfo {volume} {46}},\ \bibinfo {pages} {18678} (\bibinfo {year}
  {2021})}\BibitemShut {NoStop}%
\bibitem [{\citenamefont {et~al.}()}]{Falstad}%
  \BibitemOpen
  \bibfield  {author} {\bibinfo {author} {\bibfnamefont {B.~H.}\ \bibnamefont
  {et~al.}},\ }\href@noop {} {\bibinfo {title} {3-d quantum harmonic oszillator
  applet}},\ \bibinfo {howpublished}
  {\url{http://www.falstad.com/mathphysics.html}},\ \bibinfo {note} {retrieved
  5/30/2020}\BibitemShut {NoStop}%
\bibitem [{\citenamefont {Joffre}()}]{Quantum_online}%
  \BibitemOpen
  \bibfield  {author} {\bibinfo {author} {\bibfnamefont {M.}~\bibnamefont
  {Joffre}},\ }\href@noop {} {\bibinfo {title} {Quantum physics online}},\
  \bibinfo {howpublished} {\url{http://www.quantum-physics.polytechnique.fr}},\
  \bibinfo {note} {retrieved 5/30/2020}\BibitemShut {NoStop}%
\bibitem [{\citenamefont {Christian}\ \emph {et~al.}(2015)\citenamefont
  {Christian}, \citenamefont {Belloni}, \citenamefont {Esquembre},
  \citenamefont {Mason}, \citenamefont {Barbato},\ and\ \citenamefont
  {Riggsbee}}]{christian2015physlet}%
  \BibitemOpen
  \bibfield  {author} {\bibinfo {author} {\bibfnamefont {W.}~\bibnamefont
  {Christian}}, \bibinfo {author} {\bibfnamefont {M.}~\bibnamefont {Belloni}},
  \bibinfo {author} {\bibfnamefont {F.}~\bibnamefont {Esquembre}}, \bibinfo
  {author} {\bibfnamefont {B.~A.}\ \bibnamefont {Mason}}, \bibinfo {author}
  {\bibfnamefont {L.}~\bibnamefont {Barbato}},\ and\ \bibinfo {author}
  {\bibfnamefont {M.}~\bibnamefont {Riggsbee}},\ }\href@noop {} {\bibfield
  {journal} {\bibinfo  {journal} {The Physics Teacher}\ }\textbf {\bibinfo
  {volume} {53}},\ \bibinfo {pages} {419} (\bibinfo {year} {2015})}\BibitemShut
  {NoStop}%
\bibitem [{\citenamefont {Sayer}\ \emph {et~al.}(2017)\citenamefont {Sayer},
  \citenamefont {Maries},\ and\ \citenamefont {Singh}}]{sayer2017quantum}%
  \BibitemOpen
  \bibfield  {author} {\bibinfo {author} {\bibfnamefont {R.}~\bibnamefont
  {Sayer}}, \bibinfo {author} {\bibfnamefont {A.}~\bibnamefont {Maries}},\ and\
  \bibinfo {author} {\bibfnamefont {C.}~\bibnamefont {Singh}},\ }\href@noop {}
  {\bibfield  {journal} {\bibinfo  {journal} {Physical Review Physics Education
  Research}\ }\textbf {\bibinfo {volume} {13}},\ \bibinfo {pages} {010123}
  (\bibinfo {year} {2017})}\BibitemShut {NoStop}%
\bibitem [{\citenamefont {McKagan}\ \emph
  {et~al.}(2008{\natexlab{b}})\citenamefont {McKagan}, \citenamefont {Perkins},
  \citenamefont {Dubson}, \citenamefont {Malley}, \citenamefont {Reid},
  \citenamefont {LeMaster},\ and\ \citenamefont
  {Wieman}}]{mckagan2008developing}%
  \BibitemOpen
  \bibfield  {author} {\bibinfo {author} {\bibfnamefont {S.}~\bibnamefont
  {McKagan}}, \bibinfo {author} {\bibfnamefont {K.~K.}\ \bibnamefont
  {Perkins}}, \bibinfo {author} {\bibfnamefont {M.}~\bibnamefont {Dubson}},
  \bibinfo {author} {\bibfnamefont {C.}~\bibnamefont {Malley}}, \bibinfo
  {author} {\bibfnamefont {S.}~\bibnamefont {Reid}}, \bibinfo {author}
  {\bibfnamefont {R.}~\bibnamefont {LeMaster}},\ and\ \bibinfo {author}
  {\bibfnamefont {C.}~\bibnamefont {Wieman}},\ }\href@noop {} {\bibfield
  {journal} {\bibinfo  {journal} {American Journal of Physics}\ }\textbf
  {\bibinfo {volume} {76}},\ \bibinfo {pages} {406} (\bibinfo {year}
  {2008}{\natexlab{b}})}\BibitemShut {NoStop}%
\bibitem [{\citenamefont {Zhu}\ and\ \citenamefont
  {Singh}(2011)}]{zhu2011improving}%
  \BibitemOpen
  \bibfield  {author} {\bibinfo {author} {\bibfnamefont {G.}~\bibnamefont
  {Zhu}}\ and\ \bibinfo {author} {\bibfnamefont {C.}~\bibnamefont {Singh}},\
  }\href@noop {} {\bibfield  {journal} {\bibinfo  {journal} {American Journal
  of Physics}\ }\textbf {\bibinfo {volume} {79}},\ \bibinfo {pages} {499}
  (\bibinfo {year} {2011})}\BibitemShut {NoStop}%
\bibitem [{\citenamefont {Justice}\ \emph {et~al.}(2019)\citenamefont
  {Justice}, \citenamefont {Marshman},\ and\ \citenamefont
  {Singh}}]{justice2019improving}%
  \BibitemOpen
  \bibfield  {author} {\bibinfo {author} {\bibfnamefont {P.}~\bibnamefont
  {Justice}}, \bibinfo {author} {\bibfnamefont {E.}~\bibnamefont {Marshman}},\
  and\ \bibinfo {author} {\bibfnamefont {C.}~\bibnamefont {Singh}},\
  }\href@noop {} {\bibfield  {journal} {\bibinfo  {journal} {European Journal
  of Physics}\ }\textbf {\bibinfo {volume} {40}},\ \bibinfo {pages} {055702}
  (\bibinfo {year} {2019})}\BibitemShut {NoStop}%
\bibitem [{\citenamefont {Tytler}\ \emph {et~al.}(2013)\citenamefont {Tytler},
  \citenamefont {Prain}, \citenamefont {Hubber},\ and\ \citenamefont
  {Waldrip}}]{tytler2013constructing}%
  \BibitemOpen
  \bibfield  {author} {\bibinfo {author} {\bibfnamefont {R.}~\bibnamefont
  {Tytler}}, \bibinfo {author} {\bibfnamefont {V.}~\bibnamefont {Prain}},
  \bibinfo {author} {\bibfnamefont {P.}~\bibnamefont {Hubber}},\ and\ \bibinfo
  {author} {\bibfnamefont {B.}~\bibnamefont {Waldrip}},\ }\href@noop {} {\emph
  {\bibinfo {title} {Constructing representations to learn in science}}}\
  (\bibinfo  {publisher} {Springer Science \& Business Media},\ \bibinfo {year}
  {2013})\BibitemShut {NoStop}%
\bibitem [{\citenamefont {Treagust}\ \emph {et~al.}(2017)\citenamefont
  {Treagust}, \citenamefont {Duit},\ and\ \citenamefont
  {Fischer}}]{treagust2017multiple}%
  \BibitemOpen
  \bibfield  {author} {\bibinfo {author} {\bibfnamefont {D.~F.}\ \bibnamefont
  {Treagust}}, \bibinfo {author} {\bibfnamefont {R.}~\bibnamefont {Duit}},\
  and\ \bibinfo {author} {\bibfnamefont {H.~E.}\ \bibnamefont {Fischer}},\
  }\href@noop {} {\emph {\bibinfo {title} {Multiple representations in physics
  education}}},\ Vol.~\bibinfo {volume} {10}\ (\bibinfo  {publisher}
  {Springer},\ \bibinfo {year} {2017})\BibitemShut {NoStop}%
\bibitem [{\citenamefont {Dooren}\ \emph {et~al.}(2010)\citenamefont {Dooren},
  \citenamefont {Bock},\ and\ \citenamefont
  {Verschaffel}}]{dooren2010addition}%
  \BibitemOpen
  \bibfield  {author} {\bibinfo {author} {\bibfnamefont {W.~V.}\ \bibnamefont
  {Dooren}}, \bibinfo {author} {\bibfnamefont {D.~D.}\ \bibnamefont {Bock}},\
  and\ \bibinfo {author} {\bibfnamefont {L.}~\bibnamefont {Verschaffel}},\
  }\href@noop {} {\bibfield  {journal} {\bibinfo  {journal} {Cognition and
  Instruction}\ }\textbf {\bibinfo {volume} {28}},\ \bibinfo {pages} {360}
  (\bibinfo {year} {2010})}\BibitemShut {NoStop}%
\bibitem [{\citenamefont {disessa}\ \emph {et~al.}(2004)\citenamefont
  {disessa}, \citenamefont {Gillespie},\ and\ \citenamefont
  {Esterly}}]{disessa2004coherence}%
  \BibitemOpen
  \bibfield  {author} {\bibinfo {author} {\bibfnamefont {A.~A.}\ \bibnamefont
  {disessa}}, \bibinfo {author} {\bibfnamefont {N.~M.}\ \bibnamefont
  {Gillespie}},\ and\ \bibinfo {author} {\bibfnamefont {J.~B.}\ \bibnamefont
  {Esterly}},\ }\href@noop {} {\bibfield  {journal} {\bibinfo  {journal}
  {Cognitive science}\ }\textbf {\bibinfo {volume} {28}},\ \bibinfo {pages}
  {843} (\bibinfo {year} {2004})}\BibitemShut {NoStop}%
\bibitem [{\citenamefont {Ainsworth}(2006)}]{ainsworth2006deft}%
  \BibitemOpen
  \bibfield  {author} {\bibinfo {author} {\bibfnamefont {S.}~\bibnamefont
  {Ainsworth}},\ }\href@noop {} {\bibfield  {journal} {\bibinfo  {journal}
  {Learning and instruction}\ }\textbf {\bibinfo {volume} {16}},\ \bibinfo
  {pages} {183} (\bibinfo {year} {2006})}\BibitemShut {NoStop}%
\bibitem [{\citenamefont {Swaak}\ \emph {et~al.}(1998)\citenamefont {Swaak},
  \citenamefont {Van~Joolingen},\ and\ \citenamefont
  {De~Jong}}]{swaak1998supporting}%
  \BibitemOpen
  \bibfield  {author} {\bibinfo {author} {\bibfnamefont {J.}~\bibnamefont
  {Swaak}}, \bibinfo {author} {\bibfnamefont {W.~R.}\ \bibnamefont
  {Van~Joolingen}},\ and\ \bibinfo {author} {\bibfnamefont {T.}~\bibnamefont
  {De~Jong}},\ }\href@noop {} {\bibfield  {journal} {\bibinfo  {journal}
  {Learning and instruction}\ }\textbf {\bibinfo {volume} {8}},\ \bibinfo
  {pages} {235} (\bibinfo {year} {1998})}\BibitemShut {NoStop}%
\bibitem [{\citenamefont {Nieminen}\ \emph {et~al.}(2010)\citenamefont
  {Nieminen}, \citenamefont {Savinainen},\ and\ \citenamefont
  {Viiri}}]{nieminen2010force}%
  \BibitemOpen
  \bibfield  {author} {\bibinfo {author} {\bibfnamefont {P.}~\bibnamefont
  {Nieminen}}, \bibinfo {author} {\bibfnamefont {A.}~\bibnamefont
  {Savinainen}},\ and\ \bibinfo {author} {\bibfnamefont {J.}~\bibnamefont
  {Viiri}},\ }\href@noop {} {\bibfield  {journal} {\bibinfo  {journal}
  {Physical Review Special Topics-Physics Education Research}\ }\textbf
  {\bibinfo {volume} {6}},\ \bibinfo {pages} {020109} (\bibinfo {year}
  {2010})}\BibitemShut {NoStop}%
\bibitem [{\citenamefont {Horz}\ \emph {et~al.}(2009)\citenamefont {Horz},
  \citenamefont {Winter},\ and\ \citenamefont {Fries}}]{horz2009differential}%
  \BibitemOpen
  \bibfield  {author} {\bibinfo {author} {\bibfnamefont {H.}~\bibnamefont
  {Horz}}, \bibinfo {author} {\bibfnamefont {C.}~\bibnamefont {Winter}},\ and\
  \bibinfo {author} {\bibfnamefont {S.}~\bibnamefont {Fries}},\ }\href@noop {}
  {\bibfield  {journal} {\bibinfo  {journal} {Computers in Human Behavior}\
  }\textbf {\bibinfo {volume} {25}},\ \bibinfo {pages} {818} (\bibinfo {year}
  {2009})}\BibitemShut {NoStop}%
\bibitem [{\citenamefont {Wawro}\ \emph {et~al.}(2020)\citenamefont {Wawro},
  \citenamefont {Watson},\ and\ \citenamefont
  {Christensen}}]{wawro2020students}%
  \BibitemOpen
  \bibfield  {author} {\bibinfo {author} {\bibfnamefont {M.}~\bibnamefont
  {Wawro}}, \bibinfo {author} {\bibfnamefont {K.}~\bibnamefont {Watson}},\ and\
  \bibinfo {author} {\bibfnamefont {W.}~\bibnamefont {Christensen}},\
  }\href@noop {} {\bibfield  {journal} {\bibinfo  {journal} {Physical Review
  Physics Education Research}\ }\textbf {\bibinfo {volume} {16}},\ \bibinfo
  {pages} {020112} (\bibinfo {year} {2020})}\BibitemShut {NoStop}%
\bibitem [{\citenamefont {Just}\ and\ \citenamefont
  {Carpenter}(1976)}]{just1976eye}%
  \BibitemOpen
  \bibfield  {author} {\bibinfo {author} {\bibfnamefont {M.~A.}\ \bibnamefont
  {Just}}\ and\ \bibinfo {author} {\bibfnamefont {P.~A.}\ \bibnamefont
  {Carpenter}},\ }\href@noop {} {\bibfield  {journal} {\bibinfo  {journal}
  {Cognitive psychology}\ }\textbf {\bibinfo {volume} {8}},\ \bibinfo {pages}
  {441} (\bibinfo {year} {1976})}\BibitemShut {NoStop}%
\bibitem [{\citenamefont {Simon}\ and\ \citenamefont
  {Boyer}(1974)}]{simon1974mirrors}%
  \BibitemOpen
  \bibfield  {author} {\bibinfo {author} {\bibfnamefont {A.}~\bibnamefont
  {Simon}}\ and\ \bibinfo {author} {\bibfnamefont {E.~G.}\ \bibnamefont
  {Boyer}}\ }(\bibinfo {organization} {ERIC},\ \bibinfo {year}
  {1974})\BibitemShut {NoStop}%
\bibitem [{\citenamefont {Gegenfurtner}\ \emph {et~al.}(2011)\citenamefont
  {Gegenfurtner}, \citenamefont {Lehtinen},\ and\ \citenamefont
  {S{\"a}lj{\"o}}}]{gegenfurtner2011expertise}%
  \BibitemOpen
  \bibfield  {author} {\bibinfo {author} {\bibfnamefont {A.}~\bibnamefont
  {Gegenfurtner}}, \bibinfo {author} {\bibfnamefont {E.}~\bibnamefont
  {Lehtinen}},\ and\ \bibinfo {author} {\bibfnamefont {R.}~\bibnamefont
  {S{\"a}lj{\"o}}},\ }\href@noop {} {\bibfield  {journal} {\bibinfo  {journal}
  {Educational psychology review}\ }\textbf {\bibinfo {volume} {23}},\ \bibinfo
  {pages} {523} (\bibinfo {year} {2011})}\BibitemShut {NoStop}%
\bibitem [{\citenamefont {K{\"u}chemann}\ \emph
  {et~al.}(2020{\natexlab{b}})\citenamefont {K{\"u}chemann}, \citenamefont
  {Klein}, \citenamefont {Becker}, \citenamefont {Kumari},\ and\ \citenamefont
  {Kuhn}}]{kuchemann2020classification}%
  \BibitemOpen
  \bibfield  {author} {\bibinfo {author} {\bibfnamefont {S.}~\bibnamefont
  {K{\"u}chemann}}, \bibinfo {author} {\bibfnamefont {P.}~\bibnamefont
  {Klein}}, \bibinfo {author} {\bibfnamefont {S.}~\bibnamefont {Becker}},
  \bibinfo {author} {\bibfnamefont {N.}~\bibnamefont {Kumari}},\ and\ \bibinfo
  {author} {\bibfnamefont {J.}~\bibnamefont {Kuhn}},\ }in\ \href@noop {} {\emph
  {\bibinfo {booktitle} {CSEDU (1)}}}\ (\bibinfo {year} {2020})\ pp.\ \bibinfo
  {pages} {36--46}\BibitemShut {NoStop}%
\bibitem [{\citenamefont {Dzsotjan}\ \emph {et~al.}(2021)\citenamefont
  {Dzsotjan}, \citenamefont {Ludwig-Petsch}, \citenamefont {Mukhametov},
  \citenamefont {Ishimaru}, \citenamefont {Kuechemann},\ and\ \citenamefont
  {Kuhn}}]{dzsotjan2021predictive}%
  \BibitemOpen
  \bibfield  {author} {\bibinfo {author} {\bibfnamefont {D.}~\bibnamefont
  {Dzsotjan}}, \bibinfo {author} {\bibfnamefont {K.}~\bibnamefont
  {Ludwig-Petsch}}, \bibinfo {author} {\bibfnamefont {S.}~\bibnamefont
  {Mukhametov}}, \bibinfo {author} {\bibfnamefont {S.}~\bibnamefont
  {Ishimaru}}, \bibinfo {author} {\bibfnamefont {S.}~\bibnamefont
  {Kuechemann}},\ and\ \bibinfo {author} {\bibfnamefont {J.}~\bibnamefont
  {Kuhn}},\ }in\ \href@noop {} {\emph {\bibinfo {booktitle} {Adjunct
  Proceedings of the 2021 ACM International Joint Conference on Pervasive and
  Ubiquitous Computing and Proceedings of the 2021 ACM International Symposium
  on Wearable Computers}}}\ (\bibinfo {year} {2021})\ pp.\ \bibinfo {pages}
  {467--471}\BibitemShut {NoStop}%
\bibitem [{\citenamefont {K{\"u}chemann}\ \emph
  {et~al.}(2020{\natexlab{c}})\citenamefont {K{\"u}chemann}, \citenamefont
  {Becker}, \citenamefont {Klein},\ and\ \citenamefont
  {Kuhn}}]{kuchemann2020gaze}%
  \BibitemOpen
  \bibfield  {author} {\bibinfo {author} {\bibfnamefont {S.}~\bibnamefont
  {K{\"u}chemann}}, \bibinfo {author} {\bibfnamefont {S.}~\bibnamefont
  {Becker}}, \bibinfo {author} {\bibfnamefont {P.}~\bibnamefont {Klein}},\ and\
  \bibinfo {author} {\bibfnamefont {J.}~\bibnamefont {Kuhn}},\ }in\ \href@noop
  {} {\emph {\bibinfo {booktitle} {International Conference on Computer
  Supported Education}}}\ (\bibinfo {organization} {Springer},\ \bibinfo {year}
  {2020})\ pp.\ \bibinfo {pages} {450--467}\BibitemShut {NoStop}%
\bibitem [{\citenamefont {Andr{\'a}}\ \emph {et~al.}(2015)\citenamefont
  {Andr{\'a}}, \citenamefont {Lindstr{\"o}m}, \citenamefont {Arzarello},
  \citenamefont {Holmqvist}, \citenamefont {Robutti},\ and\ \citenamefont
  {Sabena}}]{andra2015reading}%
  \BibitemOpen
  \bibfield  {author} {\bibinfo {author} {\bibfnamefont {C.}~\bibnamefont
  {Andr{\'a}}}, \bibinfo {author} {\bibfnamefont {P.}~\bibnamefont
  {Lindstr{\"o}m}}, \bibinfo {author} {\bibfnamefont {F.}~\bibnamefont
  {Arzarello}}, \bibinfo {author} {\bibfnamefont {K.}~\bibnamefont
  {Holmqvist}}, \bibinfo {author} {\bibfnamefont {O.}~\bibnamefont {Robutti}},\
  and\ \bibinfo {author} {\bibfnamefont {C.}~\bibnamefont {Sabena}},\
  }\href@noop {} {\bibfield  {journal} {\bibinfo  {journal} {International
  Journal of Science and Mathematics Education}\ }\textbf {\bibinfo {volume}
  {13}},\ \bibinfo {pages} {237} (\bibinfo {year} {2015})}\BibitemShut
  {NoStop}%
\bibitem [{\citenamefont {Susac}\ \emph {et~al.}(2018)\citenamefont {Susac},
  \citenamefont {Bubic}, \citenamefont {Kazotti}, \citenamefont {Planinic},\
  and\ \citenamefont {Palmovic}}]{susac2018student}%
  \BibitemOpen
  \bibfield  {author} {\bibinfo {author} {\bibfnamefont {A.}~\bibnamefont
  {Susac}}, \bibinfo {author} {\bibfnamefont {A.}~\bibnamefont {Bubic}},
  \bibinfo {author} {\bibfnamefont {E.}~\bibnamefont {Kazotti}}, \bibinfo
  {author} {\bibfnamefont {M.}~\bibnamefont {Planinic}},\ and\ \bibinfo
  {author} {\bibfnamefont {M.}~\bibnamefont {Palmovic}},\ }\href@noop {}
  {\bibfield  {journal} {\bibinfo  {journal} {Physical Review Physics Education
  Research}\ }\textbf {\bibinfo {volume} {14}},\ \bibinfo {pages} {020109}
  (\bibinfo {year} {2018})}\BibitemShut {NoStop}%
\bibitem [{\citenamefont {Br{\"u}ckner}\ \emph {et~al.}(2020)\citenamefont
  {Br{\"u}ckner}, \citenamefont {Schneider}, \citenamefont
  {Zlatkin-Troitschanskaia},\ and\ \citenamefont
  {Drachsler}}]{bruckner2020epistemic}%
  \BibitemOpen
  \bibfield  {author} {\bibinfo {author} {\bibfnamefont {S.}~\bibnamefont
  {Br{\"u}ckner}}, \bibinfo {author} {\bibfnamefont {J.}~\bibnamefont
  {Schneider}}, \bibinfo {author} {\bibfnamefont {O.}~\bibnamefont
  {Zlatkin-Troitschanskaia}},\ and\ \bibinfo {author} {\bibfnamefont
  {H.}~\bibnamefont {Drachsler}},\ }\href@noop {} {\bibfield  {journal}
  {\bibinfo  {journal} {Sensors}\ }\textbf {\bibinfo {volume} {20}},\ \bibinfo
  {pages} {6908} (\bibinfo {year} {2020})}\BibitemShut {NoStop}%
\bibitem [{\citenamefont {McKagan}\ \emph {et~al.}(2010)\citenamefont
  {McKagan}, \citenamefont {Perkins},\ and\ \citenamefont
  {Wieman}}]{mckagan2010design}%
  \BibitemOpen
  \bibfield  {author} {\bibinfo {author} {\bibfnamefont {S.}~\bibnamefont
  {McKagan}}, \bibinfo {author} {\bibfnamefont {K.}~\bibnamefont {Perkins}},\
  and\ \bibinfo {author} {\bibfnamefont {C.}~\bibnamefont {Wieman}},\
  }\href@noop {} {\bibfield  {journal} {\bibinfo  {journal} {Physical Review
  Special Topics-Physics Education Research}\ }\textbf {\bibinfo {volume}
  {6}},\ \bibinfo {pages} {020121} (\bibinfo {year} {2010})}\BibitemShut
  {NoStop}%
\bibitem [{\citenamefont {Sadaghiani}\ and\ \citenamefont
  {Pollock}(2015)}]{sadaghiani2015quantum}%
  \BibitemOpen
  \bibfield  {author} {\bibinfo {author} {\bibfnamefont {H.~R.}\ \bibnamefont
  {Sadaghiani}}\ and\ \bibinfo {author} {\bibfnamefont {S.~J.}\ \bibnamefont
  {Pollock}},\ }\href@noop {} {\bibfield  {journal} {\bibinfo  {journal}
  {Physical Review Special Topics-Physics Education Research}\ }\textbf
  {\bibinfo {volume} {11}},\ \bibinfo {pages} {010110} (\bibinfo {year}
  {2015})}\BibitemShut {NoStop}%
\bibitem [{\citenamefont {Zhu}\ and\ \citenamefont
  {Singh}(2012)}]{zhu2012surveying}%
  \BibitemOpen
  \bibfield  {author} {\bibinfo {author} {\bibfnamefont {G.}~\bibnamefont
  {Zhu}}\ and\ \bibinfo {author} {\bibfnamefont {C.}~\bibnamefont {Singh}},\
  }\href@noop {} {\bibfield  {journal} {\bibinfo  {journal} {American Journal
  of Physics}\ }\textbf {\bibinfo {volume} {80}},\ \bibinfo {pages} {252}
  (\bibinfo {year} {2012})}\BibitemShut {NoStop}%
\bibitem [{\citenamefont {Landis}\ and\ \citenamefont
  {Koch}(1977)}]{landis1977measurement}%
  \BibitemOpen
  \bibfield  {author} {\bibinfo {author} {\bibfnamefont {J.~R.}\ \bibnamefont
  {Landis}}\ and\ \bibinfo {author} {\bibfnamefont {G.~G.}\ \bibnamefont
  {Koch}},\ }\href@noop {} {\bibfield  {journal} {\bibinfo  {journal}
  {biometrics}\ ,\ \bibinfo {pages} {159}} (\bibinfo {year}
  {1977})}\BibitemShut {NoStop}%
\bibitem [{\citenamefont {Hayes}(2017)}]{hayes2017introduction}%
  \BibitemOpen
  \bibfield  {author} {\bibinfo {author} {\bibfnamefont {A.~F.}\ \bibnamefont
  {Hayes}},\ }\href@noop {} {\emph {\bibinfo {title} {Introduction to
  mediation, moderation, and conditional process analysis: A regression-based
  approach}}}\ (\bibinfo  {publisher} {Guilford publications},\ \bibinfo {year}
  {2017})\BibitemShut {NoStop}%
\bibitem [{\citenamefont {Rau}\ and\ \citenamefont
  {Matthews}(2017)}]{rau2017make}%
  \BibitemOpen
  \bibfield  {author} {\bibinfo {author} {\bibfnamefont {M.~A.}\ \bibnamefont
  {Rau}}\ and\ \bibinfo {author} {\bibfnamefont {P.~G.}\ \bibnamefont
  {Matthews}},\ }\href@noop {} {\bibfield  {journal} {\bibinfo  {journal}
  {ZDM}\ }\textbf {\bibinfo {volume} {49}},\ \bibinfo {pages} {531} (\bibinfo
  {year} {2017})}\BibitemShut {NoStop}%
\bibitem [{\citenamefont {Gegenfurtner}\ \emph {et~al.}(2017)\citenamefont
  {Gegenfurtner}, \citenamefont {Lehtinen}, \citenamefont {Jarodzka},\ and\
  \citenamefont {S{\"a}lj{\"o}}}]{gegenfurtner2017effects}%
  \BibitemOpen
  \bibfield  {author} {\bibinfo {author} {\bibfnamefont {A.}~\bibnamefont
  {Gegenfurtner}}, \bibinfo {author} {\bibfnamefont {E.}~\bibnamefont
  {Lehtinen}}, \bibinfo {author} {\bibfnamefont {H.}~\bibnamefont {Jarodzka}},\
  and\ \bibinfo {author} {\bibfnamefont {R.}~\bibnamefont {S{\"a}lj{\"o}}},\
  }\href@noop {} {\bibfield  {journal} {\bibinfo  {journal} {Computers \&
  Education}\ }\textbf {\bibinfo {volume} {113}},\ \bibinfo {pages} {212}
  (\bibinfo {year} {2017})}\BibitemShut {NoStop}%
\bibitem [{\citenamefont {Dogusoy-Taylan}\ and\ \citenamefont
  {Cagiltay}(2014)}]{dogusoy2014cognitive}%
  \BibitemOpen
  \bibfield  {author} {\bibinfo {author} {\bibfnamefont {B.}~\bibnamefont
  {Dogusoy-Taylan}}\ and\ \bibinfo {author} {\bibfnamefont {K.}~\bibnamefont
  {Cagiltay}},\ }\href@noop {} {\bibfield  {journal} {\bibinfo  {journal}
  {Computers in human behavior}\ }\textbf {\bibinfo {volume} {36}},\ \bibinfo
  {pages} {82} (\bibinfo {year} {2014})}\BibitemShut {NoStop}%
\bibitem [{\citenamefont {Okan}\ \emph {et~al.}(2016)\citenamefont {Okan},
  \citenamefont {Galesic},\ and\ \citenamefont
  {Garcia-Retamero}}]{okan2016people}%
  \BibitemOpen
  \bibfield  {author} {\bibinfo {author} {\bibfnamefont {Y.}~\bibnamefont
  {Okan}}, \bibinfo {author} {\bibfnamefont {M.}~\bibnamefont {Galesic}},\ and\
  \bibinfo {author} {\bibfnamefont {R.}~\bibnamefont {Garcia-Retamero}},\
  }\href@noop {} {\bibfield  {journal} {\bibinfo  {journal} {Journal of
  Behavioral Decision Making}\ }\textbf {\bibinfo {volume} {29}},\ \bibinfo
  {pages} {271} (\bibinfo {year} {2016})}\BibitemShut {NoStop}%
\bibitem [{\citenamefont {Huang}\ and\ \citenamefont
  {Chen}(2016)}]{huang2016gender}%
  \BibitemOpen
  \bibfield  {author} {\bibinfo {author} {\bibfnamefont {P.-S.}\ \bibnamefont
  {Huang}}\ and\ \bibinfo {author} {\bibfnamefont {H.-C.}\ \bibnamefont
  {Chen}},\ }\href@noop {} {\bibfield  {journal} {\bibinfo  {journal}
  {International Journal of Science and Mathematics Education}\ }\textbf
  {\bibinfo {volume} {14}},\ \bibinfo {pages} {327} (\bibinfo {year}
  {2016})}\BibitemShut {NoStop}%
\bibitem [{\citenamefont {Ho}\ \emph {et~al.}(2014)\citenamefont {Ho},
  \citenamefont {Tsai}, \citenamefont {Wang},\ and\ \citenamefont
  {Tsai}}]{ho2014prior}%
  \BibitemOpen
  \bibfield  {author} {\bibinfo {author} {\bibfnamefont {H.~N.~J.}\
  \bibnamefont {Ho}}, \bibinfo {author} {\bibfnamefont {M.-J.}\ \bibnamefont
  {Tsai}}, \bibinfo {author} {\bibfnamefont {C.-Y.}\ \bibnamefont {Wang}},\
  and\ \bibinfo {author} {\bibfnamefont {C.-C.}\ \bibnamefont {Tsai}},\
  }\href@noop {} {\bibfield  {journal} {\bibinfo  {journal} {International
  journal of science and mathematics education}\ }\textbf {\bibinfo {volume}
  {12}},\ \bibinfo {pages} {525} (\bibinfo {year} {2014})}\BibitemShut
  {NoStop}%
\bibitem [{\citenamefont {Richter}\ \emph {et~al.}(2021)\citenamefont
  {Richter}, \citenamefont {Wehrle},\ and\ \citenamefont
  {Scheiter}}]{richter2021poor}%
  \BibitemOpen
  \bibfield  {author} {\bibinfo {author} {\bibfnamefont {J.}~\bibnamefont
  {Richter}}, \bibinfo {author} {\bibfnamefont {A.}~\bibnamefont {Wehrle}},\
  and\ \bibinfo {author} {\bibfnamefont {K.}~\bibnamefont {Scheiter}},\
  }\href@noop {} {\bibfield  {journal} {\bibinfo  {journal} {Applied Cognitive
  Psychology}\ }\textbf {\bibinfo {volume} {35}},\ \bibinfo {pages} {632}
  (\bibinfo {year} {2021})}\BibitemShut {NoStop}%
\end{thebibliography}%

\end{document}